\documentclass[twocolumn]{aastex61}
\pdfoutput=1 %for arXiv submission
\usepackage{amsmath,amstext}
\usepackage[T1]{fontenc}
\usepackage{apjfonts} 
\usepackage[figure,figure*]{hypcap}
\usepackage{url}
\usepackage{hyperref}
\usepackage{natbib}

\shorttitle{Planet Occurrence Rates using ABC}
\shortauthors{Hsu et al.}

\begin{document}

\title{Improving the Accuracy of Planet Occurrence Rates from Kepler using Approximate Bayesian Computation}
\author{Danley C. Hsu}
\affiliation{Department of Astronomy \& Astrophysics, 525 Davey Laboratory, The Pennsylvania State University, University Park, PA, 16802, USA}
\affiliation{Center for Exoplanets and Habitable Worlds, 525 Davey Laboratory, The Pennsylvania State University, University Park, PA, 16802, USA}
\affiliation{Center for Astrostatistics, 525 Davey Laboratory, The Pennsylvania State University, University Park, PA, 16802, USA}
\affiliation{Institute for CyberScience, The Pennsylvania State University}
\author{Eric B. Ford}
\affiliation{Department of Astronomy \& Astrophysics, 525 Davey Laboratory, The Pennsylvania State University, University Park, PA, 16802, USA}
\affiliation{Center for Exoplanets and Habitable Worlds, 525 Davey Laboratory, The Pennsylvania State University, University Park, PA, 16802, USA}
\affiliation{Center for Astrostatistics, 525 Davey Laboratory, The Pennsylvania State University, University Park, PA, 16802, USA}
\affiliation{Institute for CyberScience, The Pennsylvania State University}
\author{Darin Ragozzine}
\affiliation{Department of Physics \& Astronomy, N283 ESC, Brigham Young University, Provo, UT 84602, USA}
\affiliation{Department of Physics \& Space Science, 150 West University, Florida Institute of Technology, Melbourne, FL 32901, USA}
\author{Robert C. Morehead}
\affiliation{Texas Tech University, Physics \& Astronomy Department, Box 41051, Lubbock, TX 79409, USA}

\begin{abstract}
We present a new framework to characterize the occurrence rates of planet candidates identified by Kepler
based on hierarchical Bayesian modeling, Approximate Bayesian Computing (ABC), and sequential importance sampling.  
For this study we adopt a simple 2-D grid in planet radius and orbital period as our model and apply our algorithm to estimate 
occurrence rates for Q1-Q16 planet candidates orbiting around solar-type stars.
We arrive at significantly increased planet occurrence rates for small planet candidates ($R_p<1.25 R_{\oplus}$) at larger orbital periods ($P>80$d) compared to the rates estimated by the more common inverse detection efficiency method. 
Our improved methodology estimates that the occurrence rate density of small planet candidates in the habitable zone of solar-type stars is $1.6^{+1.2}_{-0.5}$ per factor of 2 in planet radius and orbital period.
Additionally, we observe a local minimum in the occurrence rate for strong planet candidates marginalized over orbital period between 1.5 and 2$R_{\oplus}$ that is consistent with previous studies. 
For future improvements, the forward modeling approach of ABC is ideally suited to incorporating multiple populations, such as planets, astrophysical false positives and pipeline false alarms, to provide accurate planet occurrence rates and uncertainties.  Furthermore, ABC provides a practical statistical framework for answering complex questions (e.g., frequency of different planetary architectures) and providing sound uncertainties, even in the face of complex selection effects, observational biases, and follow-up strategies.  
In summary, ABC offers a powerful tool for accurately characterizing a wide variety of astrophysical populations.
\end{abstract}

\keywords{methods: data analysis --- methods: statistical --- catalogs ---
planetary systems --- stars: statistics}

\section{Introduction}
Since the first exoplanets were discovered around the pulsar PSR B1257+12 \citep{WF1992},
the number of exoplanets has increased to a few thousand.  
The exoplanet population includes planet sizes, masses, and orbital properties that are not found among the Solar System planets.  
For example, there are classes of exoplanets for which there are no Solar System counterparts, including hot Jupiters, warm Neptunes, and super-Earth-size planets.  
Further, many exoplanets have highly eccentric orbits, in contrast to our Solar System in which most planets follow nearly circular orbits.  
The discovery and confirmation of such planets and planetary systems has inspired theoretical research in planet formation, planetary migration and planet-planet interactions. An essential constraint for such studies is the true underlying occurrence rate (or frequency) of exoplanets as a function of their physical and orbital properties.

\subsection{Kepler Results}
NASA's Kepler mission was launched in 2009 with the primary goal of characterizing the occurrence rate of Earth-size planets around Sun-like stars \citep{B2016,BKB+2010}.
Over the course of four years Kepler observed $\sim 192,000$ stars and identified several thousand exoplanet candidates \citep{B2014}, including a significant majority of high-quality exoplanet candidates.  
These discoveries indicate that sub-Neptune size planets represent the majority of the exoplanet population among stars surveyed and likely the Milky Way galaxy \citep{BRB+2013,BBM+2014,CMT+2016}.

While Kepler has identified a large number of exoplanet candidates, translating the Kepler planet candidate catalog into a true underlying population is a challenging.  
The observed catalog differs from the true population of exoplanets due to a variety of factors, including: 1) the geometric transit probability which depends primarily on the orbital period and star size, as well as the orbit shape, 2) the detection probability which depends primarily on the integrated transit signal-to-noise, and thus indirectly on several factors such as the transit depth, transit duration, number of transits, Kepler's photometric measurement precision stellar, and stellar photometric variability, and 3) the choice of stars targeted by Kepler.

Transit surveys are inherently more sensitive for detecting planets with short orbital
periods due to the increased geometric transit probability and increased number of transits within a given timespan of observations.  
Transit surveys are also more sensitive for detecting larger planets for a given star, since the transit signal-to-noise is proportional to the square of the planet radius.  
Similarly, transit surveys are more sensitive to the detection of planets of a given planet-star radius ratio for stars with higher quality photometry.
While it is not always easy to separate measurement uncertainty from astrophysical photometric variability, the combined differential photometric precision (CDPP) reported by the Kepler pipeline provides a useful summary of the effective ``noise'' due to the combination of astrophysical and instrumental noise sources.
Finally, the target selection process, including both the field of view of the Kepler spacecraft and which target stars were selected for data downlink, affects the planets detected and the interpretation of occurrence rates.  
Like previous studies, this study will defer on this last issue and focus on measuring the planet occurrence rate among a subset of Kepler target stars, while accounting for geometric transit probability and detection probability.

In this study, we introduce Approximate Bayesian Computing (ABC) as a tool for overcoming the above challenges to characterize the exoplanet population in general and planet occurrence rates in particular.  Our framework is original in that it can properly account for the interaction between measurement uncertainties and the planet detection probability which has been well characterized as a function of transit signal-to-noise in previous studies \citep[e.g.,][]{CCB+2015}.
We also develop a practical algorithm for applying ABC to infer planet occurrence rates
as a function of planet size and orbital period, allowing for direct comparisons to results using previous methods.  
We verify and validate the ABC algorithm using simulated data sets and report results of applying our algorithm to a recent catalog of Kepler planet candidates.  
In order to understand the context for this research, we first review several influential planet occurrence rate studies.

\subsection{Kepler Occurrence Rate Studies}
Several groups have estimated planet occurrence rates based on various Kepler planet candidate catalogs \citep[e.g.,][]{BRB+2013,BBM+2014,SGW2015,CMT+2016}.  
At the time of \citet{Y2011} only Kepler planet candidates identified based on the first two months of Kepler data were available.  
These planet candidates spanned a relatively small range of orbital periods and sizes, so the population could be reasonably modeled by a joint power law parameterization in planet radius and orbital period or a broken, joint power law.  
While \citet{Y2011} used a simplistic planet detection efficiency model, they demonstrated an efficient methodology for computing maximum likelihood estimates of occurrence rates that could generalize once the planet detection efficiency was better characterized.

Another influential early study \citep{HMB+2012} estimated the planet occurrence rate for solar type stars at each of several bins of planets, defined in terms of a 2-D grid over planet radius and orbital period.  
This study assumed planet candidates identified from the list of Kepler objects of interest (KOIs) from \citet{BKB+2011} were true planets and attempted to ``correct'' for non-detections due to either non-transiting orbital inclinations or insufficient photometric precision via the ``inverse detection efficiency method'' (IDEM), so named by \citet{FHM2014}.  The IDEM assigns each detected planet candidate with a ``weight'' that attempts to estimate the number of exoplanets with the measured size and period that would need to be distributed among the target stars in the catalog, so as to yield one detection. 

Several subsequent occurrence rates studies have built on the IDEM.  
\citet{MGL+2012} focused on improving the stellar parameters of Kepler target stars, noting that occurrence rates estimated in previous studies were biased due to the use of the Kepler Input Catalog (KIC).
\citet{FTC+2013} improved the detection efficiency model by incorporating a linear ramp model for the transit detection probability as a function of transit signal-to-noise.  
\citet{PMH2013} developed a custom transit search pipeline focusing on a subset of favorable target stars and characterize the detection efficiency of their pipeline by injecting simulated transits into the light curves.  
\citet{DC2013,DC2015,K2013} performed occurrence rate studies similar to \citet{HMB+2012},
but focused on M dwarf stars.

In another particularly influential study, \citet{CCB+2015} estimated the planet occurrence rate in the Kepler sample for several ranges of planet radii and orbital periods using the IDEM.
The main advance of \citet{CCB+2015} was quantifying the planet detection efficiency of the Kepler pipeline by injecting simulated transit signals into the raw pixel data and reprocessing the simulated data with the Kepler pipeline.  
They performed Monte Carlo simulations to characterize the planet detection efficiency.  
They fit the empirical planet detection efficiency curve with the CDF of a $\Gamma$ distribution.
They applied their improved planet detection efficiency model by applying it to Kepler planet candidate catalog.  
They focused on FGK stars observed by Kepler over Q1-Q12 and estimated planet occurrence rates for bins spanning planet sizes of $1-2 R_{\oplus}$ and orbital periods of $0.5$ to $320$ days.  \citet{CCB+2015} also computed planet occurrence rates using two alternative planet detection efficiency models:  a ``perfect detector'' model (an error function with a transition at $7.1\sigma$, assuming that each measurement has independent white noise) and the linear ramp detection efficiency model of \citet{FTC+2013}.  

\begin{figure}
\centering
\includegraphics[scale=0.43]{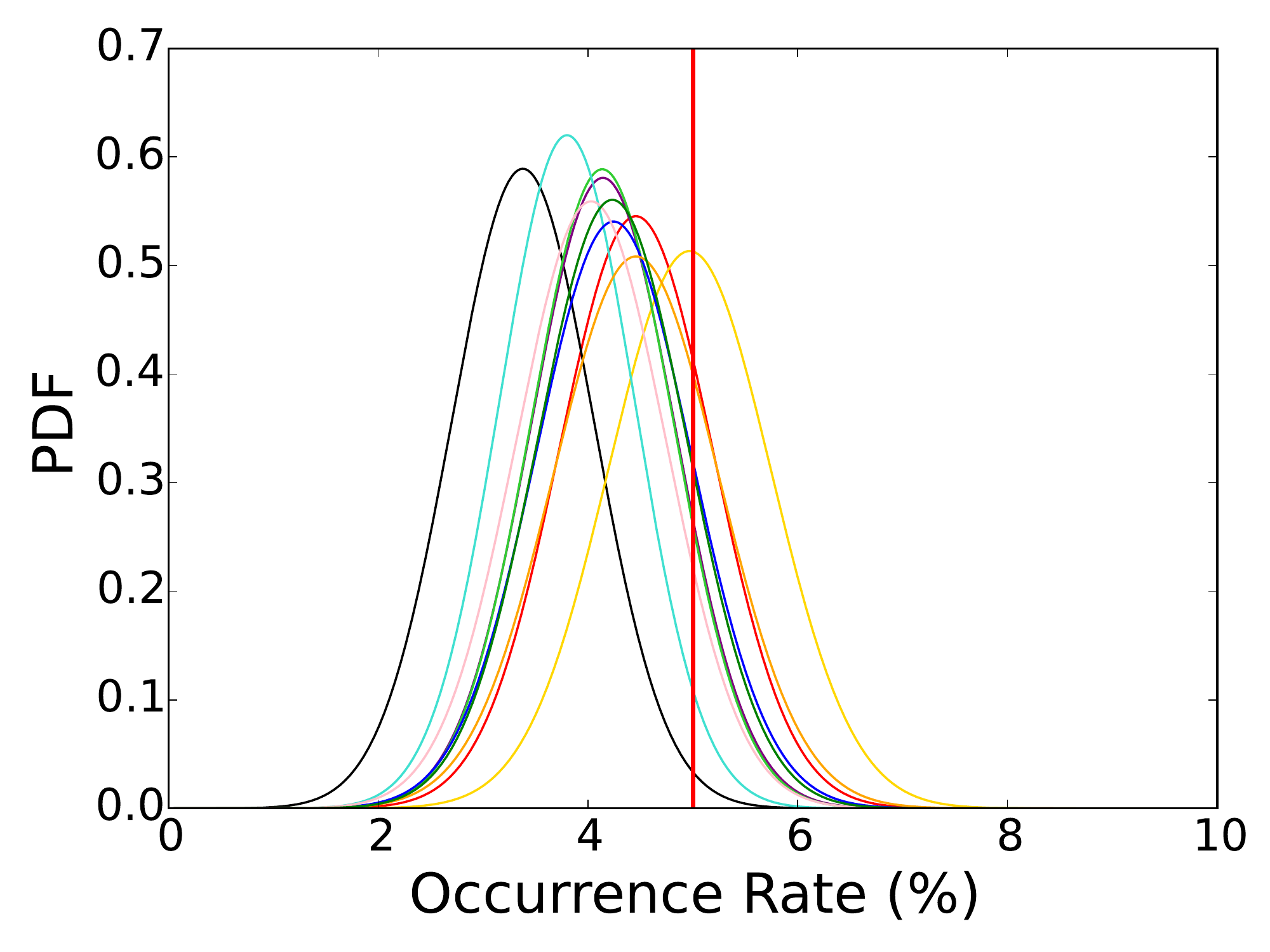}
\includegraphics[scale=0.43]{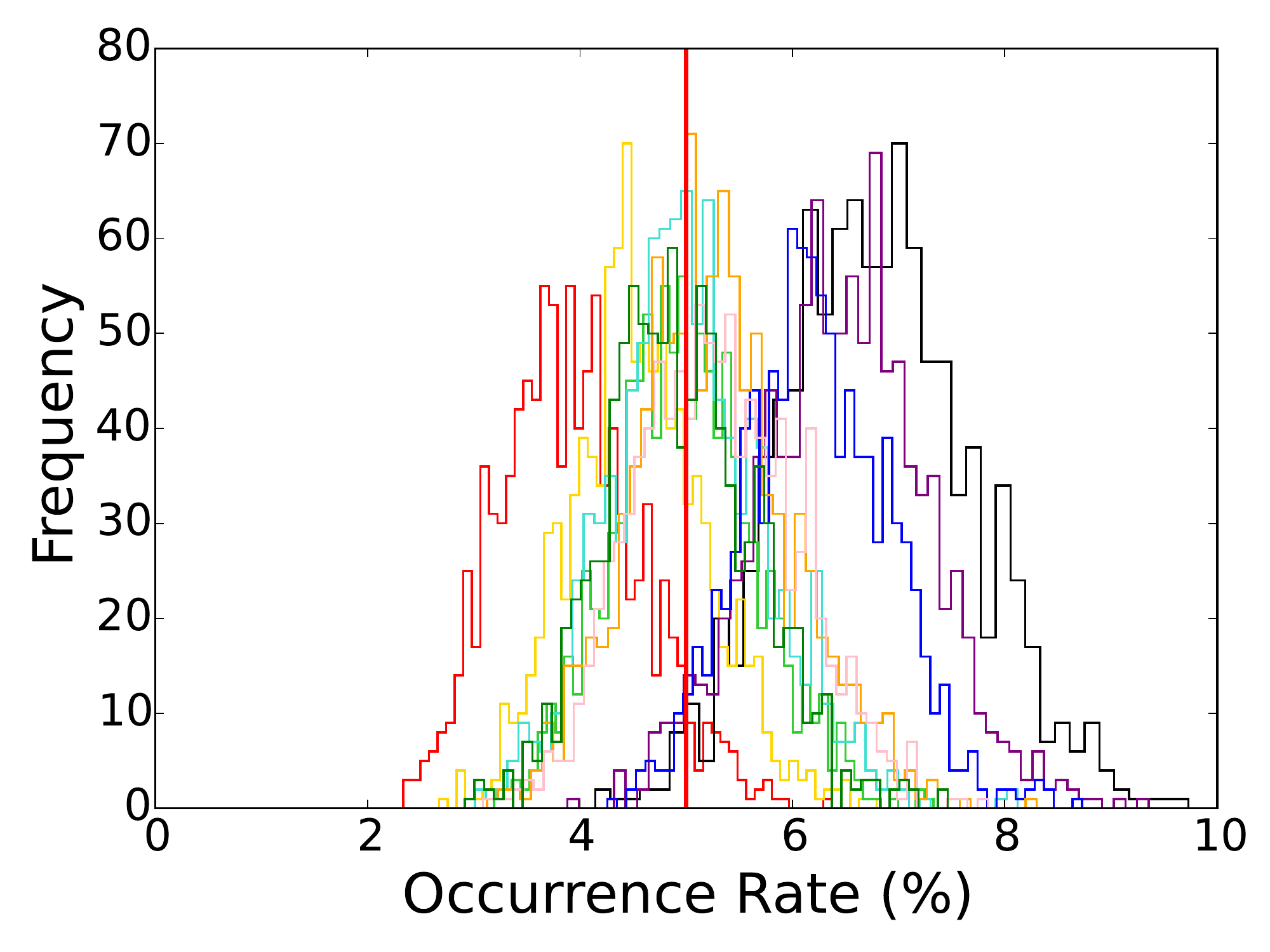}
\caption{Ten simulations where the occurrence rate is estimated for the 40-80d period, 1.25-1.5 R$_{\oplus}$ planet radius bin with a fixed ``true'' occurrence rate of 0.05 per star (indicated by the red vertical line).  
(Top) Gaussian PDF of the estimated inverse detection efficiency method occurrence rate where the width of the Gaussian $\sigma_{f}$ is defined in Appendix \ref{appIDEM}. 
The inverse detection efficiency method systematically underestimates the true occurrence rate.
(Bottom) Histogram of 1000 samples drawn from the final ABC posterior.
The ABC simulations accurately bracket the true occurrence rate.}
\label{figabc-invdet_comp}
\end{figure}

While the simplicity and speed of the IDEM is appealing, the reliance on estimated planet properties can lead to a significant bias.
To illustrate this problem, we created two sets of 10 planet catalogs defined by a $0.05$ per star occurrence rate in the $P=40-80$ day and $R_{\mathrm{p}}=1.25-1.5 R_{\oplus}$ bin (see \S\ref{secVV}). Each catalog is then treated as an observed planet candidate catalog to be used to estimate occurrence rates by the IDEM.  
In Fig. \ref{figabc-invdet_comp} (top panel) we plot the occurrence rate estimated by IDEM as a Gaussian PDF with a mean and width calculated according to Appendix \ref{appIDEM}.  
This demonstrates the IDEM estimated rates are biased and systematically underestimate the occurrence rate.  
This can be easily understood as a result of using the estimated planet size to compute the completeness correction, rather than the true planet size.  
While the resulting bias is small for most bins, it becomes substantial for planets near the threshold of detection as is the case for this bin.  
Since small planets at orbital periods near one year are both near the threshold of detection and of particular interest for both science and mission planning, it is important to develop methods that more accurately characterize planet occurrence rates of such planets.
The bottom panel of Fig. \ref{figabc-invdet_comp} shows results from an analogous computation using a more rigorous hierarchical Bayesian model and Approximate Bayesian Computing, as described in \S\ref{secStatisticalModel}.  
We will discuss these results in \S\ref{secVV} once we have described the methodology in detail.  
For now, we merely note that the ABC posteriors are correctly centered on the true occurrence rate, in stark contrast to the estimates based on the IDEM.
The bias in the IDEM becomes even more pronounced for larger orbital periods and/or smaller planets, as expected due to the large measurement uncertainties. 

\citet{FHM2014} described a more rigorous methodology for characterizing the exoplanet occurrence rates based on a hierarchical Bayesian model (HBM).  
Similar to \citet{HMB+2012} and \citet{CCB+2015}, they estimated planet occurrence rates over a 2-D grid in terms of planet radius and orbital period.  
However, unlike previous studies they assumed that the occurrence rates for nearby bins were correlated, making use of a Gaussian Process (GP) prior for the occurrence rates within each bin.
Recognizing the importance of an accurate planet detection efficiency method, they applied their methodology to a planet catalog from \citet{PMH2013}, as \citet{CCB+2015} was not yet available.  
\citet{FHM2014} also found a significantly increased rate of small, long-period planets compared to the inverse detection efficiency method used in the \citet{CCB+2015} study, due to a combination of using a HBM and their GP prior.  However, their inferred results were less than those of \citet{PMH2013} who focused their attention on a subset of target stars and used planet candidates from a custom detection pipeline.

Recently, \citet{BCM+2015} characterized the population of small exoplanets by combining a more recent Kepler planet candidate catalog including data from Q1-16 \citep{TJS+2014}, an improved planet detection efficiency model, based on planet injection studies which injected transits at the pixel level \citep{CCB+2015}, and a Bayesian generalization of \citet{Y2011}.  
Rather than using a grid in planet size and orbital period, \citet{BCM+2015} assume a broken, joint power law planet distribution function over planet radii and orbital period parameter space.  Given the limitations of their parameterization, they focus on small planets ($0.75-2.5  R_{\oplus}$) with orbital periods in the range 50-300 days. \cite{BCM+2015} neglect uncertainty in planet radius, and do not account for how uncertainty in planet radius affects the transit detection efficiency.  Their inferred rate of small planets was also higher than \citet{CCB+2015}.

In another recent paper, \citet{SGW2015} applied methods somewhat similar to those presented in this study to the planet catalog generated by the Kepler pipeline.  Here, we describe an approach for performing planet occurrence rate studies based on a forward model, while also providing a solid statistical foundation perform statistical inference.  

\subsection{Future of Kepler Occurrence Rate Studies}
Previous studies have focused on the occurrence rate of planets on an individual basis, and side-stepped complications due to multiple planet systems.  
Previous planet occurrence rates studies that included all Kepler planet candidates around a star can be better interpreted as estimating the average number of planets (within a range of sizes and orbital periods) per star, which may be quite difference from the fraction of stars with such planets \citep{Y2011,BR2016}.
Given the limitations of early Kepler planet candidate catalogs, some previous studies included only the ``first'' or ``strongest'' detected planet candidate.  These are even more difficult to interpret, since whether any given planet is counted depends on the existence of other planets in the same system that differ in size and/or orbital period.  
Thus, such studies do not estimate the fraction of stars having at least one planet within that bin \citep{BR2016}.  
In principle, a hierarchical Bayesian framework is well-suited to characterizing the occurrence rate of planetary systems.  However, a direct generalization of the \citet{FHM2014} model to multiple planet systems would be computationally prohibitive.

The primary purpose of this study is to validate a new statistical framework for inferring
planet occurrence rates that can later be generalized to address the occurrence rate of planetary systems.  
In particular, we evaluate a HBM for the exoplanet occurrence rate, using ABC and sequential importance sampling.  
This allows us to avoid the bias in planet occurrence rates introduced by the IDEM. 
We also apply our methods to a recent Kepler planet candidate catalog to provide updated planet occurrence rate estimates as a function of planet size and orbital period.   
Of course, these estimates could be improved with further, as we make several assumptions common among published planet occurrence rate studies (e.g., host star properties, catalog reliability).

To ease comparisons, our study will compare occurrence rates estimated using ABC to those of the inverse detection efficiency method using a star and planet catalog similar to that of \citet{CCB+2015} and the $\Gamma$ cumulative distribution function (CDF) detection efficiency model of \citet{CCB+2015}.  We describe our statistical methodology in \S\ref{secStatisticalModel}.  
In \S\ref{secPhysicalModel}, we provide the details of the forward model for the planet population and detection model used by ABC.  We validate our methodology on simulated data in \S\ref{secVV} and apply it to actual Kepler data in \S\ref{secApplication}.  Finally, we discuss the implications of our results and future prospects for characterizing exoplanet populations in \S\ref{secDiscussion}.

\subsection{Role of ABC}
Originally, Bayesian inference via forward modeling and rejection was intended as a pedagogical tool for understanding Bayesian inference \citep{rubin1984}.  The combination of increasing computational power and the complexity of genetic evolution models, led to ABC being considered for practical calculations \citep{Tavare1997}.  Several authors have explored the choice of summary statistics \citep[e.g.,][]{RSSB:RSSB1010,Marin2012} as well as sampling algorithms to improve the sampling efficiency of ABC.  Our sequential importance sampling algorithm (sometimes known as particle or sequential Monte Carlo in the statistics community) closely follows that of \citet{BCM+2008} and \citet{RCM+2013}.

ABC has had some limited applications in the field of astronomy 
over the past few years.  \citet{CP2012} gave one of the 
earliest demonstrations of 
ABC in astronomy by making use of a Sequential Monte-Carlo 
implementation of ABC to determine posteriors for a stochastic
model of high-redshift massive galaxy morphological transformation.
In a later study, an Markov chain Monte Carlo (ABC-MCMC) variant
of the algorithm was used to study the shape of the thick disc
of the Milky Way galaxy in order to better constrain formation
models \citep{RRF+2014}.  

Following these applications in galaxy
evolution, several cosmology papers used applications of ABC.
\citet{ARA+2015} detailed a Population Monte-Carlo implementation
of ABC (ABC-PMC) and applied it to  
calibrate image simulations for wide-field 
cosmological surveys.  \citet{IVC+2015}
released a public version of ABC-PMC written in \emph{Python} 
called COSMOABC and demonstrated it by estimating posteriors
for cosmological parameters using galaxy cluster number counts.
A new implementation of ABC called superABC is discussed by 
\citet{JWS2016} in their study wherein cosmological parameters
are constrained using Type Ia supernovae light curves with 
the use of two different distance functions.
This study applies similar algorithms to characterize the occurrence rate of exoplanets.

\section{Statistical Framework}
\label{secStatisticalModel}
Bayesian statistical analysis is becoming common in astronomy thanks to the increasing availability of practical algorithms and computing power to perform the required calculations.  
Unlike traditional means of fitting a best-fit model to data (e.g. using $\chi^{2}$), a Bayesian analysis can estimate the posterior probability distribution ($\pi (\theta|Y)$) for a given set of model parameters ($\theta$) given the data $Y$, prior information about the model parameters $\pi(\theta)$, and a likelihood function $f(Y|\theta)$ which specifies the probability of the data given the model and model parameters:
\begin{equation}
\label{eqnBayes}
\pi(\theta|Y) = \frac{\pi(\theta) f(Y|\theta)}{\int_{\theta}\,d\theta' \pi(\theta') f(Y|\theta')}.
\end{equation}
The primary challenge to widespread application of Bayesian methods is computing the potentially high-dimensional integrals involved.  
Markov chain Monte Carlo (MCMC) is a powerful tool for Bayesian parameter estimation in exoplanet research.  
Even simple implementations are often adequate for low-dimensional problems, such as characterizing the orbit of a single planet \citep{F2005}.  
In higher dimensions (e.g., characterizing multiple planet systems), efficient sampling often demands more sophisticated samplers \citep[e.g.,][]{NFP2014}.

\subsection{Hierarchical Bayesian Models}
\label{secHBM}
HBM provides a rigorous statistical framework for characterizing a population.  
In HBM, we separate model parameters into parameters that describe the population ($\phi$) and parameters that describe the properties of individual objects from that population ($\theta_j$), for $j=1...N_t$, where $N_t$ is the number of members of the population.  
For our study, $\phi$ will refer to population-level model parameters, such as the occurrence rate of planets (for a given range of planet size, orbital period and host star properties), while $\theta_j$ would refer to the properties of the $j$th target (e.g., planet size, orbital period, epoch of transit, host star properties).  
Similarly, we can divide the observed data ($Y$) into subsets ($Y_j$) that depend only on the data for the $j$th target.  
If we are interested in characterizing the population, then we marginalize over the properties of each target (i.e., integrate over these parameters, weighting by their relative probability), so Bayes' theorem can then be written as
\begin{equation}
\label{eqnHBM}
\pi(\phi|Y) = \frac{ \pi(\phi) \prod_{i=1}^{N_t} \int_{\theta_j} \, d\theta_j \pi(\theta_j|\phi) f(Y_j|\theta_j) }{ \int_{\phi'} \, d\phi' \pi(\phi')  \prod_{i=1}^{N_t}  \int_{\theta_j} \, d\theta_j \pi(\theta_j|\phi') f(Y_j|\theta_j) }.
\end{equation}
Note that $Y_j$ depends only on $\theta_j$ and is conditionally independent from the population parameters, $\phi$, or the properties of other planets.  This conditionally independence means that if we knew $\theta_j$, then we would not gain any additional information about the probability distribution for $Y_j$ by making use of knowledge about the distribution or values of $\phi$ or $\theta_{k~\ne~j}$.  
The conditional independence significantly simplifies the calculations, transforming high dimensional integrals into the product of many lower dimensional integrals.  
Another common simplification is to replace the observed data $Y$ with summary statistics that describe the key properties of the data ($s=S(Y)$).  
As a simplistic example, one might approximate an entire light curve due to a single planet transiting a photometrically quiet star as a constant out-of-transit flux, a transit depth, transit duration, orbital period, epoch of one transit and magnitude of Gaussian measurement noise.   
If one uses an accurate model for the data, then the true model parameters form sufficient summary statistics. However, the model parameters are often not directly observable.  
In practice, the true model parameters must be estimated from the data ($Y$) or a set of summary statistics.

Despite these simplifications, evaluating Eqn. \ref{eqnHBM} for either the set of all Kepler light curves or a Kepler planet candidate catalog would be extremely expensive.  
First, note that $N_t$ refers to the number of targets and not the number of planets.  
Even if we reduce the Kepler observations to merely four properties per planet (orbital period, orbital phase, transit depth, transit duration) and two properties per star (star radius, magnitude of measurement noise), evaluating the likelihood in Eqn. \ref{eqnHBM} would require integrating over $\simeq~N_t (2+4\left<N_{pps}\right>)$ dimensions, where $\left<N_{pps}\right>$ is the average number of planets per star.  
Since $N_t>100,000$ and $\left<N_{pps}\right>$ is likely greater than 2, even a simplistic model would require integrating over $\sim~10^6$ dimensions.

For targets where all the star and planet properties are well measured, the integral over $\theta_j$ could likely be performed efficiently using importance sampling and the measured planet properties.  
However, even with an ideal observatory, most planets will not be detected in a transit search due to transit geometry.  
Since many (if not most) targets harbor planets that are not detected, the integrals over $\theta_j$ need to be evaluated over all possible planet sizes and orbital periods for each of the planets.  
If each target star could harbor only a single planet, then one could treat each planet as independent to approximate this integral once for each target and reuse the result each time the likelihood needed to be evaluated \citep[e.g.,][]{Y2011,FHM2014}.  This approach would no longer be possible once we allow each host star to have multiple planets which may not be distributed independent of each other (e.g., avoid each other due to stability or correlations in orbital period).
The sizes and orbital periods of planets around a given star are not independent of each other, so the properties of any planets that are detected affect the conditional distributions for the undetected planets.  
The abundance of stars with multiple planets necessitates accounting for multiple planets around each star.  
Thus, evaluating the likelihood in Eqn.\ \ref{eqnHBM} in the context of the Kepler planet search and multiple planet systems is intractable, regardless of whether it is to be used for an MCMC simulation to perform parameter estimation for the population parameters, $\phi$, or for a computing maximum likelihood estimator.  
These limitations motivate us to consider an alternative approach to characterizing the exoplanet population that we will explore in the next section.  

\subsection{Theory of Approximate Bayesian Computing}
\label{secABC}
Approximate Bayesian Computation (ABC) provides a means to perform Bayesian inference for problems where either it is not practical to write out the likelihood or evaluating the likelihood is computationally prohibitive.  

ABC is particularly well-suited for applying HBM to model populations.
One specifies a forward model to describe both the population to be characterized and the process of collecting data (see \S\ref{secPhysicalModel}) and a process for choosing which simulated datasets (and their associated populations parameters) are to be accepted.  
One computes many realizations of the forward model via some variant of Monte Carlo simulation and accumulates an ensemble of model parameters that yielded simulated data sets that are sufficiently similar to the observed data set.  
The result is an ``ABC posterior'' ($\pi_{\mathrm{ABC}}(\phi|Y)$) that takes the place of a traditional Bayesian posterior.  
If we retain only the population level parameters, then we effectively marginalize over the unknown physical properties of individual members of the population, including both those detected and those that escaped detection:  
\begin{equation}
\label{eqnABCposterior}
\pi_{\mathrm{ABC}}(\phi|Y)  = \pi(\phi|\rho(S(Y),S(Y_{\mathrm{obs}}))<\epsilon) 
\end{equation}
where $\rho$ is the distance function and $\epsilon$ is the distance threshold.
Based on the challenges described in \S\ref{secHBM}, we anticipate that this approach will be particularly valuable for characterizing the population of planetary systems.  

While ABC is motivated by complex problems for which likelihood-based methods are impractical, one can also apply ABC to problems for which one can write down the priors and likelihood.  In such cases, one can show that the ABC posterior approaches the true posterior for small $\epsilon$ if one uses sufficient summary statistics and a reasonable distance function \citep{Marin2012,BCM+2008}. 
Therefore, ABC provides a solid statistical foundation for performing likelihood-based statistical inference for complex models where inference using the full likelihood is impractical.

Here we will provide an outline of the general process for choosing which simulated data sets are to be accepted, before addressing the specific implementation for this study in subsequent sections.  
First, we choose a set of summary statistics ($S(Y)$) that characterize the key properties of the observed or simulated data (see \S\ref{secSummaryStats}).  
Second, we choose a distance function, $\rho(S(Y_i),S(Y_{\mathrm{obs}}))$, that specifies the distance between the summary statistics computed for a simulated dataset ($s_i=S(Y_i)$) and the summary statistics for the observed data, ($s_{\mathrm{obs}}=S(Y_{\mathrm{obs}})$; see \ref{secDistance}).
Third, we specify $\epsilon$, the maximum acceptable distance for a simulated data set to be accepted.   
In this case, the ABC posterior is given by 
\begin{equation}
\label{eqnHBMABC}
\begin{split}
\pi_{\mathrm{ABC}}(\phi|Y) \propto & \\
& \int_{\phi} \, d\phi \; \delta(\rho(S(Y),S(Y_{\mathrm{obs}}))<\epsilon) \times \\
& \prod_{i=1}^{N_t} \int_{\theta_j} \, d\theta_j \, \pi(\phi) \pi(\theta_j|\phi) f(Y_j|\theta_j) 
\end{split}
\end{equation}
The outer integral samples from the prior for $\phi$, the inner integrals correspond to sampling the possible values for the true parameters of each population member ($\theta_j$), and the  $\delta$-function selects those draws that result in the distance function meeting the acceptance criterion.

As a particularly simple example, one could use a naive Monte Carlo algorithm, repeatedly drawing sets of model parameters $(\phi_i$,$\theta_{ij})$ from the prior distributions $\pi (\phi_i) \sim \pi(\phi)$ and $\theta_{ij} \sim \pi(\theta_j|\phi_i$), where ``$\sim$'' can be read as ``is distributed as.''  
For each realization, one would generate a simulated data set ($Y_i$) and compute the associated summary statistics, $s_i = S(Y_i)$.  
If the distance between the simulated data and the observed data is sufficiently small, i.e., $\rho(s_{i},s_{\rm{obs}}) \le \epsilon$, then the model parameters $\phi_i$ and $\theta_{ij}$'s are added to the ensemble that forms the ABC posterior. In other words, only parameter values that create simulated data that are very close to the observed data are kept in the ABC posterior. 

In practice, the naive ABC algorithm described above is extremely computationally inefficient for most problems of interest and small $\epsilon$.  
For ABC to be useful on real world problems, one must apply a more efficient sampling strategy \citep[e.g.,][]{BCM+2008,B2009,RCM+2013}.  
For this study, we apply a sequential importance sampling strategy described in \S\ref{secABCPMC}.

\subsection{Approximate Bayesian Computing in Practice}
\label{secABCPMC}
This study makes use of an ABC - Population Monte Carlo (ABC-PMC) algorithm which relies on sequential importance sampling to evolve an ABC posterior with the goal of achieving a large number of draws with sufficiently small $\epsilon$.  
Our algorithm closely follows that described in \citet{BCM+2008} and \citet{RCM+2013}.
Each generation in the computed sequence contains an ensemble of model parameters.  
In statistical parlance, each generation's ensemble (or population) consists of multiple ``particles''.
Each ensemble of particles can be used to approximate a probability distribution over the model parameter space.
To initialize the ABC-PMC algorithm, we first apply simple Monte Carlo with a fixed number of draws from the prior distribution, and we select the $N_{\mathrm{part}}$ particles which resulted in the smallest distances to the observed dataset.  
The selected particles make up the 0th generation in which all particles have a distance less than $\epsilon_0 = \max_i(\rho_{0,i})$.  
For this and each subsequent generation, we construct a Gaussian mixture model with the mixture components centered on the model parameter values of each particle.  
For the 0th generation, each particle is assigned an equal mixture weight which will be updated on subsequent generations.  
Each mixture component shares a common covariance matrix which is based on the sample covariance of the current particle ensemble.
Mathematically, the ABC posterior at generation $g$ ($p_{\mathrm{ABC},g}(\phi | Y_{\mathrm{obs}} )$) is estimated by
\begin{equation}
p_{\mathrm{ABC},g}(\phi | Y_{\mathrm{obs}} ) = \sum_{i=1}^{N_{\mathrm{part}}} w_{g,i} N_{\phi}(\phi_{g,i}, \Sigma_{g} )
\end{equation}
where
$w_{g,i}$ is the weight of the $i$th particle in generation $g$,
the weights at each generation form a simplex (i.e., $\sum_i w_{g,i} = 1$),
$N_{\phi}(\mu,\Sigma)$ is a normal probability distribution for $\phi$ with mean $\mu$ and variance $\Sigma$, and
$\Sigma_{g}$ is the sample covariance of the population parameters $\phi_{g,i}$ at generation $g$.

For each subsequent generation, the ABC-PMC particle population is evolved using importance sampling (IS) to draw trial sets of model parameters ($\phi^*$) for the next generation, $\phi^* \sim p_{IS,g}(\phi)$.  
When evaluating trial sets of model parameters for the $g$th generation, only trials that result in a distance between the simulated data and observed data that is less than $\epsilon_g$ are accepted.  
The distance threshold $\epsilon_g$ is gradually decreased, so that each generation more closely approximates the posterior distribution (which requires $\epsilon \rightarrow 0$).  
Once the algorithm draws $N_{\mathrm{part}}$ trial particles with $\rho^{*}<\epsilon_{g}$, the weights for each particle are updated to be proportional to the ratio of the prior and the importance sampling density.  
First, we compute unnormalized weights, $w^*_{g,i} = p(\phi_{g,i}) /
p_{IS,g}(\phi_{g,i} )$, before computing the properly normalized weights, $w_{g,i} = w^*_{g,i} / \sum_{i=1}^{N_{\mathrm{part}}} w^*_{g,i}$.  
This results in the ABC posterior in subsequent generations respecting both the prior for model parameters and the distance constraint \citep{BCM+2008} .

\citet{BCM+2008} recommend using a Gaussian mixture model for the importance sampling density,
\begin{equation}
\phi^* \sim p_{IS,g+1}(\phi) = \sum_{i=1}^{N_{\mathrm{part}}} w_{g,i} N_{\phi}(\phi_{g,i}, \tau \Sigma_{g} ),
\end{equation}
where the covariance is scaled by $\tau\simeq~2$ relative to the ABC posterior, so as to reduce the risk of some particles being assigned large weights.  By performing this scaling, we effectively sample generation $g+1$ from an expanded posterior of generation $g$.
We found that sometimes this can result in the particle population being dominated by a small number of particles that are assigned large weights.  
To reduce this risk and make the ABC-PMC algorithm more efficient, we suggest truncating each component of the mixture model to exclude values of $\phi^*$ that result in $(\phi^* - \phi_{g,i})' (\tau \Sigma_g)^{-1} (\phi^* - \phi_{g,i}) \ge X_{\mathrm{crit}}$, where $X_{\mathrm{crit}} = 2 \sqrt{n_{\mathrm{param}}}$ and $n_{\mathrm{param}}$ is the number of model parameters being characterized by ABC.  
Formally, this can result in some valid regions of parameter space for $\phi$ being assigned zero probability by the importance sampler.  
In practice, this is not a problem for our application, since any such regions are extremely unlikely to result in simulated data set with $\rho(S(\phi^*),s_{\mathrm{obs}})\le\epsilon_{g+1}$.  
Note that truncation is not used when approximating the final ABC posterior, so the ABC posterior is non-zero for all $\phi$ (with prior support).

During the course of this study, we found that the distance threshold $\epsilon_i$ can be decreased relatively rapidly in early generations.  
Eventually, $\epsilon_i$ becomes small enough that 
sampling noise is likely to cause a simulated data set to have a distance larger than $\epsilon_i$.  
At this point, further reductions in $\epsilon_i$ rapidly become computationally prohibitive.  
Therefore, we developed a set of heuristics to automatically terminate the sequential importance sampler once the desired precision had been reached or once the algorithm efficiency has dropped to the point where it is no longer making significant improvements.  
We halt once of the following events occurs:
(1) The mean distance between the generated catalogs and the true observed catalog drops below a target distance threshold $\epsilon$,
(2) the number of generations reaches a maximum limit, $N_{\mathrm{max,gen}}=200$,
(3) the number of consecutively repeated states exceeds a threshold which we set equal to the the total number of particles. A particle's state is repeated when the algorithm fails to generate a set of model parameters resulting in $\rho(s^*,s_{\rm{obs}}) < \epsilon_{i}$ after $N_{\rm{max}}$ trial sets of model parameters.  The algorithm tracks how many times each particle has repeated its current set of parameter values, and resets the counter each time the particle is successfully updated.  If the number of consecutively repeated states summed over all particles in a generation exceeds the number of particles per generation, then the algorithm halts.
(4) during the most recent generation, the median number of trial sets of model parameters required to generate a successful set of $N_{\mathrm{part}}$ model parameters (i.e., $\rho(s^*,s_{\rm{obs}}) <\epsilon_{i}$) exceeds $0.2\times~N_{\rm{max}}$, or
(5) the algorithm fails to improve the distance threshold for three consecutive generations due to at least one particle in each generation requiring a number of attempted draws greater than $0.75~N_{\rm{max}}$.
Ideally, one would use only Criteria 1, so as to achieve the desired level of convergence.  Often, one does not know a priori what level of convergence is practical.  In such cases, one can set $\epsilon$ to be so small that criterion 1 is unlikely to be met.   In this case, the algorithm will proceed until one of criteria 2, 3, 4 or 5 are met, at which point further improvements in $\epsilon$ become extremely computationally expensive.  In our calculations, criteria 2 and 3 were never invoked.  Instead, criteria 4 and 5 was found to be effective stopping criteria, recognizing when further calculations would yield negligible improvements in $\epsilon$.  At this point, the ABC posterior width had converged to very near the the Monte-Carlo limit for the vast majority of our calculations (see Fig. \ref{eps-sig-tests}).

\subsection{Summary Statistics}
\label{secSummaryStats}
The ability of the ABC posterior to approximate the true posterior depends on using a distance function that can identify simulated data sets that are similar to the observed data set.  
In practice, the distance function $(\rho)$ is usually chosen to be a function of a set of summary statistics ($S$), rather than a function of all the data ($Y$).

For example, consider the problem of characterizing the distribution of a population of values ($x_i$), each drawn from a normal distribution with unknown mean ($\mu$) and variance of unity.  
One can compute the posterior distribution for $\mu$ using only $\left<x\right>$, the mean value of the $x_i$'s.
If one knows $\left<x\right>$, then also knowing the specific value of each of the $x_i$'s would not affect one's estimate of $\mu$ or the population from which $x_i$'s are drawn.
For this reason, $\left<x\right>$ is referred to as a ``sufficient statistic'' to characterize a normal distribution with known variance.  
When using sufficient statistics and a reasonable distance function, one can prove that the ABC posterior will approach the true Bayesian posterior for sufficiently small $\epsilon$.  
Similarly, if the true distribution were normal with unknown mean and unknown variance, the sample mean and sample covariance would form a set of sufficient statistics.  
Formally, the full data set is also a sufficient statistic, but this is rarely useful, since it is unlikely that one would draw values close to each of the $x_i$'s, even if one were using the true population mean and variance.

In this paper, we will characterize planet occurrence rates over a 2-d grid in planet size and orbital period, focusing on a specific set of $N_{\mathrm{targ}}$ host stars.  We define $n_{s,r,p}$ to be the number of planets with size $R_{p,r}<R_{p}\le R_{p,r+1}$ and orbital period $P_{p}<P\le P_{p+1}$ orbiting star $s$.  We assume that each $n_{s,r,p}$ is drawn independently from a Poisson distribution with occurrence rate $f_{r,p}$ that is constant over the relevant range of radii \& periods.  
If we assume a known number of target stars, that this model is the true distribution of planets, and that all planets are detected, then the number of planets in each bin would form a sufficient set of summary statistics.
Even if not all planets are detected, the detected number of planets in each bin: $n_{r,p} = \sum_s n_{s,r,p}$ are sufficient summary statistics provided that the probability of detection for each planet is known.  
Therefore, we adopt the ratio of the number of planets in each bin to the overall number of target stars ($S(Y) = \{ n_{r,p}/N_{\mathrm{targ}} \}$) as our summary statistics.  
Of course, it is usually necessary to perform statistical inference with a model whose assumptions are not exactly true.  We discuss some of the more obvious issues related to the choice of summary statistics below and return to discuss the issue of model misspecification further in \S\ref{secModelMisspecification}.

We chose bin boundaries of \{0.5, 1.25, 2.5, 5, 10, 20, 40, 80, 160, 320\} days \& \{0.5, 0.75, 1, 1.25, 1.5, 1.75, 2, 2.5, 3, 4, 6, 8, 12, 16\} $R_{\oplus}$, following \citet{CCB+2015} and \citet{HMB+2012}.  Of course, future applications could make alternative choices for the boundaries of planet size and orbital period bins.  If one were to adopt bins that were very large, then the assumption of a constant occurrence rate within the bin could become inadequate.  If one were to adopt bins that were very small, then most bins would have zero to a few planets, causing the variance of the posterior occurrence rate to be large.  
In principle, one could adopt a prior that assumes occurrence rates in nearby bins are correlate \citep{FHM2014} to reduce the variance while retaining significant model flexibility.  In this study, we assume rates are uncorrelated between bins, so as to simplify the interpretation and comparisons to previous studies.  

Since orbital periods are measured precisely, there is essentially no error in assigning planets to period bins for any reasonable bin width.  
Uncertainty in transit depths and stellar radii propagate to cause a modest difference between the true and estimated planet radii, resulting in some planets near a boundary in planet size being assigned to the ``wrong'' radius bin.  ABC can accounts for this effect, by using a forward model that generates both true and observed stellar and planetary radii.  The main consequence of uncertainty in planet size is that any sharp changes in the true occurrence rate as a function of planet radius will appear to be less sharp in the observed (and simulated) catalogs. 
A more subtle effect of the uncertainty in stellar radii is that the posterior for the planet occurrence rate in neighboring bins in planet size could be anti-correlated.  If a planet were near the boundary of two bins in planet size, then it could be assigned to one bin or the other, but not both bins or neither bin.  If one asked what is the occurrence rate for a range of planet sizes spanned by both bins, then the sum of the occurrence rates would be accurate, but the uncertainty in the sum of the occurrence rates for the two bins would not be as large is if the two bins were uncorrelated.  We investigate the strength of these effects in \S\ref{secMultiBinResults}.

For more complex problems (e.g., characterizing planetary architectures), it is likely not possible to identify a small set of sufficient summary statistics.  
Yet, the power of ABC can be traced to the dimensional reduction that comes from a relatively small set of summary statistics encapsulating the key features of the data.   
Just as a standard Bayesian analysis must choose an appropriate prior and likelihood, in an ABC analysis, it is important to choose an appropriate set of summary statistics, so as to encode the key physical properties of the population for the scientific questions at hand.  

\subsection{Distance Functions}
\label{secDistance}
The ABC algorithm requires specifying a function to compute the distance ($\rho$) between the summary statistics for observed catalog $s_{\mathrm{obs}}$ and the summary statistics for each simulated planet population ($s^*$).  
For the simulations shown in this paper, we consider the planet occurrence rate for each bin of planet size and orbital period separately.  When characterizing the occurrence rate of a single bin, the summary statistic is a scalar, so the choice of the distance function is obvious: 
$\rho((s_{\mathrm{obs}},s^*) = \left(s_{\mathrm{obs}} - s^*\right)^2$.  
In other words, the trial parameters ($\phi^{*}$) will be accepted in the ABC posterior
for generation $g$ if the result of the forward model produces a frequency of detected planets
($n_{r,p}/N_{\mathrm{targ}}$) that is within $\sqrt{\epsilon_{g}}$ of the Kepler observed frequency $s_{\mathrm{obs}}$.

In the primary applications for this paper, each bin is independent of the other bins.  However, to illustrate how ABC could be applied to more complex models with multiple interdependent summary statistics, we also explore distance functions that simultaneously fit occurrence rates in multiple different bins.
In simulations that consider the occurrence rate for multiple bins simultaneously, the summary statistics are no longer scalar.  
In preliminary simulations, we explored various choices for the distance function, including:
1) $\rho_{\max}(s_{\mathrm{obs}},s^*) = \max_k |s_{\mathrm{obs},k} - s^*_k|$,
2) $\rho_{L1}(s_{\mathrm{obs}},s^*) = \sum_k |s_{\mathrm{obs}} - s^*_k|$, and
3) $\rho_{L2}(s_{\mathrm{obs}},s^*) = \sum_k |s_{\mathrm{obs}} - s^*_k|^2$,
where the maximum or sum over $k$ is over each of the summary statistics (i.e., over all the bins). 

When analyzing models that include multiple bins with widely varying values of $f_{r,p}$, we found that $\rho_{L1}$ or $\rho_{L2}$ could result in less precise estimates of some $f_{r,p}$'s.  
However, when considering occurrence rates for a large 
number of bins simultaneously (i.e. $>5$ bins) $\rho_{\max}$ 
struggles significantly in the efficiency of 
convergence.
Due to this poor efficiency for the $\rho_{\max}$ distance function, we choose to report results based on the $\rho_{L2}$ distance function in this study. 
In this study, each of the summary statistics is a sample occurrence rate.  
In future studies where summary statistics are not naturally comparable, it will likely be advantageous to normalize or scale the different summary statistics.  

Our implementation of the ABC-PMC algorithm described above is provided by the ABC.jl package \citep{ABC_Julia} for the \href{http://julialang.org/}{Julia programming language} \citep{Julia} and is available at \url{https://github.com/eford/ABC.jl/tree/hsu_etal_2018-v1.0} under the MIT Expat License.
ABC.jl was designed to be generally applicable and is not specific to our current applications of characterizing exoplanet populations.

\subsection{Model Misspecification}
\label{secModelMisspecification}

One should be concerned about the potential effects of model misspecification when applying any statistical model to a real physical situation.  In the case of a traditional Bayesian analysis, this concern is manifest in the choice of the likelihood function.  In the case of ABC, concerns about model misspecification can appear in either the forward model or the choice of summary statistics.  While the potential for model misspecification likely renders the choice of summary statistics imperfect, the underlying concerns of model misspecification are not unique to ABC, but apply to any statistical analysis, whether frequentist or Bayesian, with or without a likelihood.

Dealing with model misspecification is a major problem well beyond the scope of this paper.  We will discuss two examples of how our model is likely imperfect below.
First, we note that either ABC or a hierarchical Bayesian model are less sensitive to deviations of the true and assumed planet detection efficiency model than the IDEM method, since IDEM uses a point estimate for the planet detection efficiency.  Further, the bias in IDEM persists, even when planet detection efficiency is known exactly.
Using ABC offers the benefit of a rigorous statistical foundation for performing inference even in presence of the inevitable uncertainty in the planet size.

The distribution of planets likely differs from the assumed piecewise constant model (or any other possible assumed distribution).  
Simple parametric planet occurrence rate models (e.g., power-law) are prone to bias due to model misspecification, particularly if the empirical constraints are significantly stronger in one region of parameter space (e.g., where planet detections are more abundant) than another region where one would like to characterize the predictive distribution (e.g., Earth-size planets in habitable zone).  A non-parametric model for the planet occurrence rate provides increased flexibility to reduce such risks of introducing a bias in the occurrence rate of Earth-analogs due to model misspecification.  Of course, the specific choice of number and location of bin boundaries is subjective.  We discuss the associated tradeoffs in \S\ref{secSummaryStats} and adopt bin boundaries following \citet{CCB+2015} and \citet{HMB+2012}.  
 
Similarly, the planet detection efficiency model is not known perfectly.  Just as a standard frequentist or Bayesian analysis using a likelihood based on an imperfect detection model would be biased, the results of ABC would be biased by an inaccurate detection probability model.  
Fortunately, the magnitude of this bias goes to zero as the assumed planet detection efficiency model approaches the true planet detection efficiency.  
The Kepler team has performed numerous simulations to characterize the detection probability as a function of the injected planet parameters.
We apply a detection probability for each of the detected planets based on Monte Carlo simulations from transit injection studies \citep[e.g.,][]{CCB+2015}.  
This dramatically mitigates the concern about the accuracy of the planet detection efficiency model, particularly relative to analyses which assume a detection efficiency model without the benefit of calibration to Monte Carlo simulations.  Future studies can explore the sensitivity of results to the choice of planet detection efficiency model.

\section{Physical Model: SysSim}
\label{secPhysicalModel}
Our forward model has two key components: a process for generating a set of target stars and associated planets, and a model for ``observing'' the targets to obtain a catalog of detected planets and their estimated properties.  We describe the model for generating target stars and planets in this section and the model for generating simulated observed catalogs in \S\ref{secObservedCatalog}.
We refer to both parts of the forward model as SysSim, short for the Planetary Systems Simulator.  SysSim is an empirical model which generates planetary systems using
flexible parameters (here $f_{r,p}$) that can be later compared to planet formation hypotheses.
Its primary goal is to aid in interpreting Kepler data to characterize the true, underlying exoplanetary population.
SysSim is implemented in Julia \citep{Julia} by the ExoplanetsSysSim.jl package \citep{ExoplanetsSysSim} and is available at
\url{https://github.com/dch216/ExoplanetsSysSim.jl/tree/hsu_etal_2018-v1.0} 
under the MIT Expat License.
SysSim was designed to be highly extensible and also contains many additional features, some tested and some still under development, which we intend to present in future publications.

\subsection{Stellar Properties}
\label{secStarProperties}
For each simulated planet catalog, we draw a sample of $N_{\mathrm{targ}}=150,518$ target stars from the Q1-Q16 catalog of Kepler target stars \citep{HSM+2014}.  
For each target star, we store the estimated star mass, estimated star radius, the robust root mean square combined differential photometric precision on 4.5 hour timescale (CDPP), the timespan from the first to the last Kepler observation of the target, and the duty cycle (i.e., the fraction of above time span during which usable Kepler data is available), as well as upper and lower uncertainty estimates for the star mass and radius.

For this study, we populate the catalog of potential target stars with a subset of the Kepler stellar properties catalog, as obtained from the Exoplanet Archive at NExScI in June 2015, which derive from \citet{HSM+2014}.  
Following \citet{CCB+2015}, we aim to include main sequence FGK stars by including only target stars that:
1) were observed for at least one quarter by Kepler in Q1-Q12,
2) have estimated properties of $4000$K$\ <T_{\rm{eff}}<7000$K and $\log g > 4.0$, and
3) have valid values and uncertainties for star mass, radius, density, and CDPP.
Our resulting target star list is very similar to that of \citet{CCB+2015}, but differs slightly, since some of the data in the star properties catalog at NExScI has been updated since the study.  
For example, we find 150,518 stars, which is slightly less than the 152,066 found by \citet{CCB+2015}.

Our approach for assigning stellar properties differs from that of assigning planet properties (see \S\ref{secPlanetModel}), since we are not attempting to infer properties of the Kepler target star population.  In principle, one could try to simultaneously infer the properties of the Kepler host star population and the planet population.  However, this would require substantially increasing the number of model parameters and observational constraints, as well as modeling the complex target selection process for choosing which targets would be downloaded.  Future studies may find it advantageous to model both the host star and planet populations.  This could be particularly useful when applied to results from the upcoming TESS mission.  Since the TESS mission plans to download full frame images, the target selection process is significantly simplified relative to that of the Kepler planet search.

\subsection{Planet Model}
\label{secPlanetModel}
Each planet is assigned physical properties  (i.e., radius, period, orbit) by drawing from proposed distributions for a given model.  
The sampling distribution for each planet is set by the model parameters and is not related to measurements of any specific planet candidate in the Kepler catalog.  Planet properties are drawn independently, even for planets orbiting the same star.
For each target star, we draw the number of planets from a Poisson distribution with rate, $\lambda = \sum_{r,p} f_{r,p}$, where $r$ and $p$ range over all planet radius and orbital period bins being included in the simulation at hand.  
For this paper, the summary statistics depend only on the total number of planets in each bin.  Since planets are drawn independently of each other, we could have used a simpler model that did not allow for multiple planets around the same star.  
Nevertheless, the code implements this feature in preparation for future studies with SysSim that will study the architectures of planetary systems. 
Note that $\lambda$ is the average number of planets per star (in the entire radius/period range).  
Experimentation showed that large $\lambda$ would slow the analysis and was 
inconsistent with the observations, so the Poisson distribution is truncated at $N_{\mathrm{max,pl}} = 10$.  
For each planet, we first assign a radius-period bin by drawing from a categorical distribution with the probability for each bin proportional to the rate $f_{r,p}$ and normalized so the probabilities sum to unity.  
Next, we draw the precise orbital period and planet size, uniformly in log period and log radius constrained to be within the respective ranges of planet's assigned bin.
Each planet is assigned a Keplerian orbit, where the eccentricity is drawn from a Rayleigh distribution with scale parameter $\sigma_{hk}=0.03$ \citep{FLR+2014}.   Such small eccentricities have only a modest effect on the planet occurrence rate \citep{BCM+2015}, but we include full Keplerian orbital parameters for the sake of completeness and future applications.   
Next, inclinations are drawn so that orbital planes are oriented isotropically over the sky (i.e. $\cos i\sim U(-1,1)$).  Since we are considering only the rate of planets per star (per bin in period and radius) and not the properties of multiple planet systems, the distribution of mutual inclinations of planets does not affect our calculations.  

The angles (argument of periastron, longitude of ascending node, and mean anomaly at the reference epoch) are drawn from uniform distributions.  
Collectively, each simulated catalog of target stars and their planets is known as a ``physical catalog'', since this contains all the stars and planets (including those that will not be detectable), along with their true properties (as opposed to their estimated properties).

\subsection{Observed Catalog}
\label{secObservedCatalog}
For each simulated physical catalog, we construct a simulated ``observed catalog'' that will be compared to the actual catalog of Kepler planet candidates.  
Rather than generating full light curves for each target, we compute the primary observables that characterize each transiting planet: the orbital period ($P$), the time of the $0$th transit ($t_{0}$), the fractional transit depth ($d=R_{p}/R_{\star}$), and the transit duration ($D$).

Next, we compute the probability that each planet would be detected.  
In practice, the detection probability depends on two parts: the geometric probability that the planet transits the star for a given observer and the probability that the planet would be detected if it were known to transit the star.  
For the sake of computational efficiency, we first compute the probability that the planet would be detected if the orbital plane resulted in the planet transiting across the diameter of the star.  
For this study, we we adopt the $\Gamma$ distribution CDF model for the planet detection efficiency ($p_\mathrm{det}$) which accounts for the transit depth, duration, CDPP, orbital period, timespan of observations and duty cycle \citep{CCB+2015}.  
We multiply the above detection probability by a binomial analytic window function to ensure that only planets transiting at least three times are considered detectable ($p_{\mathrm{win,}\ge3}$),  following \citep{BCM+2015}.  
At this point, planets with negligible detection probability (even if observed edge-on) are discarded from the observed catalog.  
The remaining planets are considered ``potentially detectable'', meaning there is a non-negligible probability that the planet would be detected if the viewing geometry were favorable.

Next, we account for the fact that not all planets transit their star for a given viewing geometry.  At this point, there are two possible approaches:  adopting a single viewing geometry or averaging over all possible viewing geometries.   
The simplest approach is to compute which of the potentially detectable planets actually transit the star given their orbits and star properties for a single observer (i.e., using the specific $i$ and $\omega$ of each planet).  
In this case, we simply test whether $a\cos(i) (1-e^2) \le R_\star (1+e\sin\omega)$, which requires that the center of the planet pass inside the disk of the star, to determine whether the geometric transit probability $p_\mathrm{geo}$ for this specific planet and viewing geometry is unity or zero.  
In our ABC simulations, the stellar radius $R_\star$ used for calculating the transit probability is the physical stellar radius which is assigned based on the observed radius and its uncertainty reported in the input stellar catalog.  
(We ignore the small numbers of planets with impact parameters greater than unity, since such transits are rare and are so hard to distinguish from grazing eclipsing binaries that such a transit would often be dismissed as a likely astrophysical false positive.)
For each simulated potentially detectable planet that transits the disk of its host star, we compute an updated transit detection probability that now accounts for its actual impact parameter which typically results in reducing the transit duration and the integrated transit signal-to-noise.
Each potentially detectable planet that transits its host star is labeled as either detected or not detected by drawing from a Bernoulli distribution with the updated transit detection probability.  
This results in a combined probability of detecting a planet, $p_{\mathrm{comb}} = p_\mathrm{geo} p_\mathrm{det} p_{\mathrm{win},\ge3}$).

Each detected planet is added to an observed catalog, along with of its measured transit parameters, referring to the measured orbital period ($\hat{P}$), measured time of the $0$th transit ($\hat{t}_{0}$), measured fractional transit depth ($\hat{d}$), and measured transit duration ($\hat{D}$).
Each measurement is assumed to be normally distributed about the true value with dispersion based on the transit signal-to-noise and uncertainty in stellar parameters.
For simplicity, we use only the diagonal terms of the covariance matrix for the transit parameters, as given by Eqn. A8 and Table 1 of \citet{PR2014}. 
Then, planets are assigned to their radius-period bin based on their observed radius and period, which are calculated using the measured transit parameters in the observed planet and stellar catalogs.  If a planet's observed radius falls
outside the range of radii under consideration, then it is excluded when counting the number of planets detected in the relevant bin(s).
The summary statistics are simply the number of detected planets in each planet radius-period bin divided by the number of target stars within the catalog, $\{ n_{r,p}/N_{\mathrm{targ}} \}$.  

An alternative approach is to generate a probabilistic catalog, where each potentially detectable planet is included along with a weight proportional to its combined probability.  
Future studies characterizing the distribution of planetary architectures could benefit from CORBITS \citep{BR2016} which can efficiently compute the sky-averaged geometric probability of any combination of planets transiting  their host star using nearly algebraic formulae.
While SysSim includes support for probabilistic catalogs, this feature was not used for this study, since we focus on properties of individual planets rather than planetary systems.  
We provide a brief overview of this approach in Appendix \ref{appProbabilisticCatalog}.

\section{Verification and Validation}
\label{secVV}
Having described our statistical framework in \S\ref{secStatisticalModel} and the particulars of our physically-motivated forward model in \S\ref{secPhysicalModel}, we proceed to the task of verification and validation of our code and algorithms before applying them to actual science data.  
In this section, we verify that our implementations of ABC-PMC and SysSim, as well as our choices for the summary statistics and distance function, accurately characterize planet occurrence rates for simulated data sets.  
We also validate that our refinements to ABC described in \S\ref{secABCPMC} actually improve the algorithms' efficiency, at least for our specific problem.

\subsection{Methods}
\label{secVVMethods}
For our initial verification tests, we generated synthetic ``observed'' catalogs from our forward model that will be used in place of the actual catalog of Kepler planet candidates for the purpose of validating our algorithm \& code throughout this section.  
Both the synthetic observed catalog and catalogs simulated during the ABC process are parameterized by the planet occurrence rates ($f_{r,p}$) for population level parameters, $\phi$.

We apply ABC-PMC to generate $p_{\mathrm{ABC},\epsilon}(f_{r,p} | Y_{\mathrm{obs}} )$, a Gaussian mixture model approximation for the ABC posterior distribution with tolerance $\epsilon$ for each bin in planet radius and orbital period, indexed by $r$ and $p$ for our synthetic data $Y_{\mathrm{obs}}$.  
Since we marginalize over the physical properties of each planet, the ABC posterior is a function of only the population-level parameters, in this case $f_{r,p}$.
During the sequential importance sampling phase of ABC, we use 40 ``particles'' for our mixture model. 
We run ABC for several generations, decreasing the distance threshold until we reach the stopping criterion.
For these validation runs, we set the target distance threshold, $\epsilon$, to zero.  In practice, this resulted in ABC stopping once the median number of attempted draws per particle exceeds 20\% the maximum number of attempts per particle, $N_{\rm{max}}= 50$.

We compute the mean ($\hat{f}_{r,p}$) and a 68.3\% credible interval of the ABC posterior for each $f_{r,p}$ using the weights from importance sampling and compare this credible interval to the true simulated planet occurrence rate.  

\subsection{Results of V\&V}
\subsubsection{ABC Posteriors are Accurate}
\label{secVVResults}

We confirm that the mean of the ABC posterior provides an accurate estimate of the true occurrence rate.  
We repeat this experiment ten times for three different bins using different simulated catalogs to confirm that that 68.3\% credible interval for the occurrence rate accurately approximates the difference between the posterior mean and true occurrence rate and that there is no detectable bias (e.g., Fig.\ \ref{figabc-invdet_comp}, bottom panel).
Note that our ABC posteriors behave much better than those of the inverse detection efficiency method which can give significantly biased estimates (see Fig.\ \ref{figabc-invdet_comp}, top panel).
We provide a more detailed description of the inverse detection efficiency method in Appendix \ref{appIDEM}.

\subsubsection{Width of ABC Posterior \& Choice of Distance Tolerance}
\label{secEpsilon}
In principle, the width of the ABC posterior should be larger than width of the true posterior when the final distance tolerance is greater than zero.  If the distance function is chosen well and the ABC simulation is executed effectively, then $\epsilon$ will be sufficiently small that the width of the ABC posterior will be close to that of the true posterior.
In this section, we demonstrate that the width of ABC posteriors very closely approximates the correct posterior width by comparing results for different values of $\epsilon$ and by comparing to a simplified Bayesian model in the limit of low fixed transit noise included in the forward model for ABC (see Appendix \ref{appPoisson}).  

\begin{figure}
\centering
\includegraphics[scale=0.28]{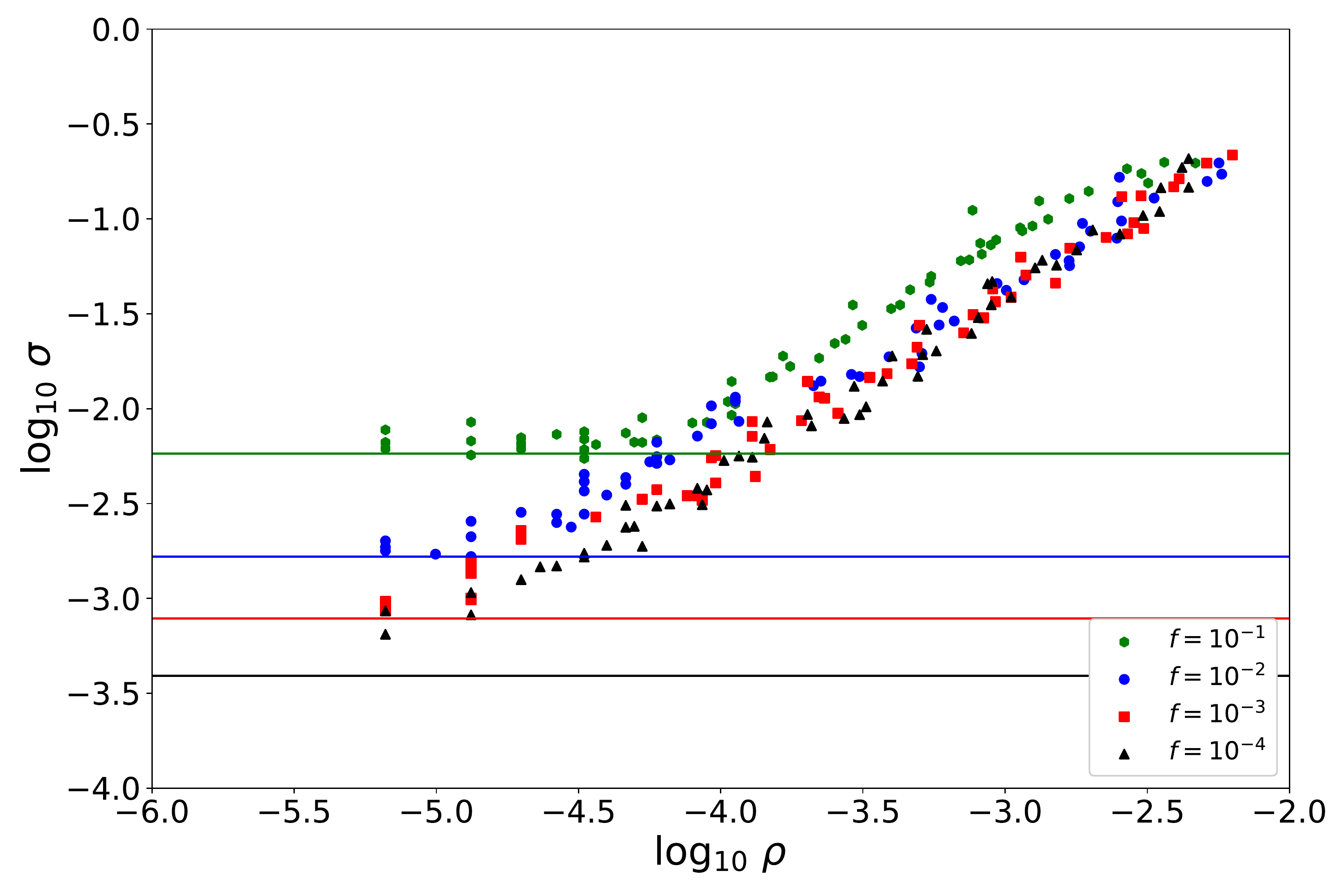}
\caption{Standard deviation, $\sigma$, of the ABC posterior for the occurrence rates as a function of $\rho$, the median distance within each generation for $P=10-20$ days and $R_{\mathrm{p}}=1-1.25 R_{\oplus}$ \& $N_{\star}= 150,518$.  
The symbol style/color indicates simulations with different underlying occurrence rates. Each simulation starts from the top right and decreases in $\sigma$ towards the bottom left over the course of the run. The colored lines correspond to the inevitable Monte Carlo uncertainty, estimated by the posterior width for simulations using the Simplified Bayesian model described in Appendix \ref{appPoisson}. The plateau of the different simulations at their respective Monte Carlo uncertainties show that our simulations reach the Monte Carlo noise limit.
}
\label{eps-sig-tests}
\end{figure}

First, we characterize how $\sigma$, the half-width of the 68.3\% credible interval for the ABC posterior for $f$, depends on the $\epsilon$.  
For this purpose, we perform tests using a single bin in planet size ($R_{\mathrm{p}} = 1-1.25 R_{\oplus}$) and period ($P = 10-20$ days).
For each simulation we plot the half-width of the ABC posterior, $\sigma_g$, as a function of $\rho_g$, the median distance of the particles in the $g$th generation of the sequential importance sampler (Fig. \ref{eps-sig-tests}).  
During early generations, $\rho_g$ steadily decreases and $\sigma$ improves, moving towards the bottom left of the figure.  
Eventually, randomness inherent in the forward model and the finite number of targets results in the ABC posterior width fluctuating, even as $\epsilon_g$ continues to decrease.  
As long as one uses an $\epsilon$ value no greater than the point where $\sigma_g$ plateaus, the ABC posterior width is insensitive to the choice of $\epsilon$.  
However the computational cost can increase significantly, if one insists on a significantly smaller value of $\epsilon$.  

Next, we compare the width of the ABC posterior to estimates for the standard deviation of the posterior from a Bayesian analysis for a simplified model which accounts for the finite sample size, but does not require ABC (solid lines).  We see in Figure \ref{eps-sig-tests} that the widths of the ABC posterior for simulations with $f = 10^{-1}$, $10^{-2}$, and $10^{-3}$ are very near the theoretical limits for the posterior width of the simplified Bayesian model that we describe in Appendix \ref{appPoisson}.
In the case of the $f=10^{-4}$ simulation, the ABC stopping criterion was met before $\sigma_g$ had clearly plateaued.
If we were to continue to run the ABC simulations with even smaller $\epsilon$, then the ABC posterior width might decrease further, but this would be computationally expensive.  
Based on comparing to the posterior width for the simplified Bayesian model, we expect that the posterior width could not decrease by more than a factor of $\sim~2$. 
It is also worth exploring why the ABC posterior width is greater than the width of the simplified Bayesian model.  
For the $f=10^{-4}$ simulation, there were 0 planets detected, so the data impose a strong limit on the planet occurrence rate, but the shape of the posterior at smaller rates will be sensitive to the choice of priors.  
Therefore, if zero planet candidates are detected in any bin, then we report only only upper limits on the planet occurrence rate.  
For example, in the case of the $f=10^{-4}$ simulation, we would report a limit of $<6\times10^{-4}$.  
In contrast, for bins with no detected planets, the IDEM method as commonly used in the literature provides neither an estimate of the rate nor an upper limit.  
Of course, it is possible to compute upper limits using either a maximum likelihood or a Bayesian analysis without using ABC (e.g., see Appendix \ref{appIDEM}).

\subsubsection{Width of ABC Posterior as a Function of Orbital Period}
\begin{figure}
\centering
\includegraphics[scale=0.28]{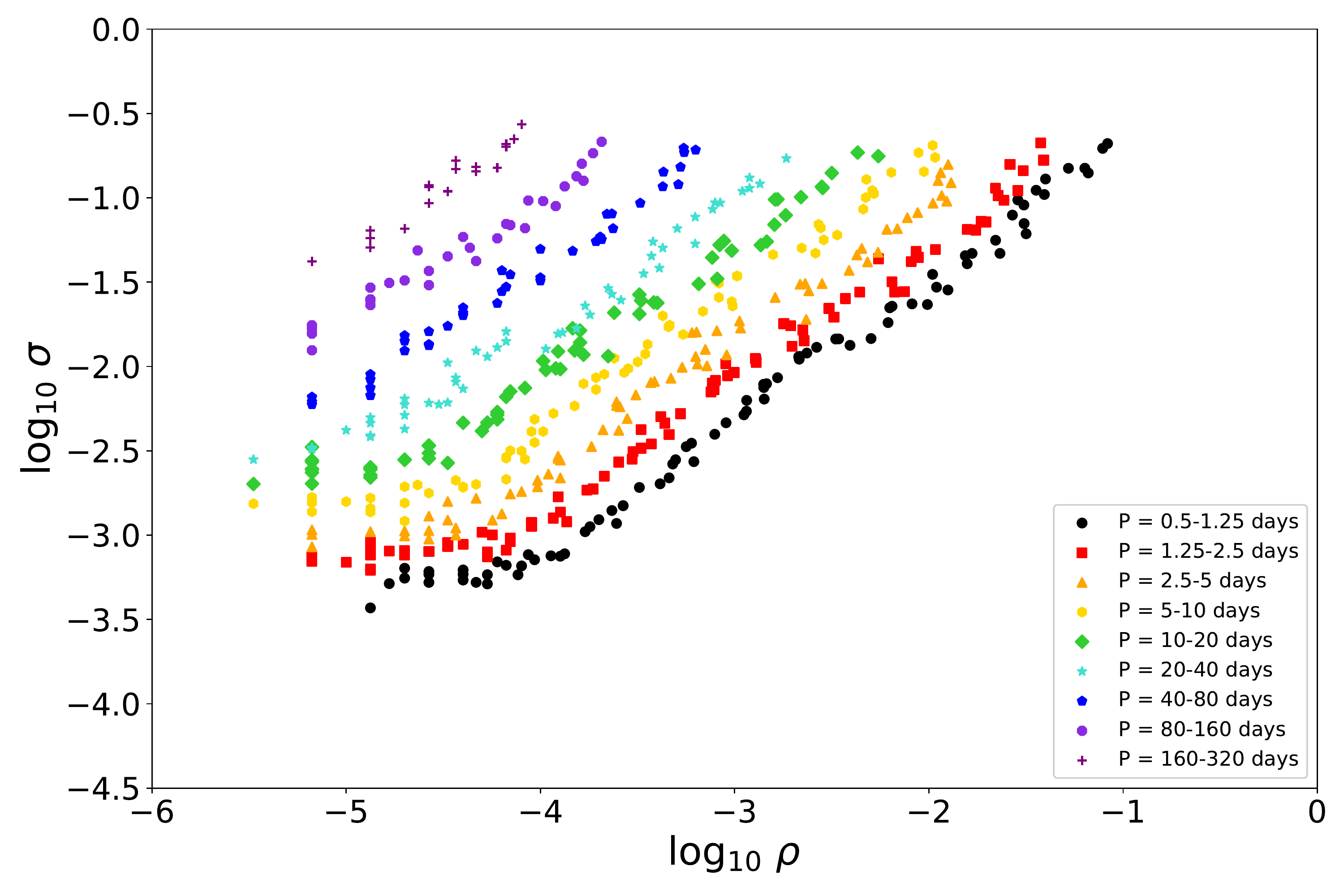}
\caption{Standard deviation, $\sigma$, of the ABC posterior for the occurrence rates as a function of $\rho$, the median distance within each generation $\epsilon$ for $N_{\star}= 150,000$.  
The true occurrence rate used to generate the observed catalog is fixed at $f = 0.015$.
The symbol style indicates simulations with bins of different orbital period ranges.  Each simulation starts from the top right and decreases in $\sigma$ towards the bottom left over the course of the run.  As expected, more precise estimate of the occurrence rate (i.e., smaller $\sigma$) are achieved at shorter orbital periods, primarily due to the increased transit probability.}
\label{figper-eps-sig}
\end{figure}

Since the transit probability decreases with increasing orbital period, there are more planets detected at small orbital periods.
Therefore, we expect the precision of planet occurrence rates will increase with orbital period for a fixed occurrence rate and the width of the ABC posteriors will increase for bins with increasing orbital period.  
To explore this effect, we perform simulations on bins over the full period range, keeping the radius range of each bin the same ($R_{\mathrm{p}} = 1-1.25 R_{\oplus}$).  
For one set of simulations, the true occurrence rate used to generate the observed catalog is fixed at $f = 0.015$ per star (Fig.\ \ref{figper-eps-sig}).
First, we note that the ABC posterior accurately estimates the true occurrence rates (not shown) for all of these cases.   
As expected, we find more precise estimates of the estimated occurrence rate (i.e., smaller values of $\sigma$) at shorter orbital periods.  
For periods less than 20 days, the width of the ABC posterior has clearly plateaued indicating that further reductions in $\epsilon$ will not affect the width of the ABC posterior.  
For orbital periods greater than 80 days, ABC has not reached a plateau, so smaller values of $\epsilon$ might result in a smaller uncertainty on planet occurrence rates.  

\subsection{Number of Model Parameters}
\label{secNParam}
For most of this study, we perform inference on a single population-level model parameter at a time.  
However, in \S\ref{secRadiusUncertainty} we report results based on simulations simultaneously estimating occurrence rates for multiple bins include differing planet size ranges, but a common orbital period.   
We interpret the results to understand the effect of uncertainty in planet radii on the joint ABC posterior for planet occurrence rates in bins of neighboring planet sizes.  

Before performing such tests, we need to verify that the ABC-PMC algorithm also performs acceptably when estimating multiple $f_{r,p}$'s simultaneously.  
Therefore, we characterize the computational efficiency of our algorithm as a function of the number of planet radius-period bins being characterized simultaneously.  
These simulations also serve as a stepping stone to future studies that include more population-level model parameters.   

For these experiments, we choose the true values of the planet occurrence rates ($f_{r,p}$) to be similar to those in the DE2 detection table in Figure 7 of \citet{CCB+2015}.
For the distance function, we use $\rho_{\mathrm{max}}$ from \S\ref{secDistance} (i.e. compute the absolute difference between the ratio of the number of detected planets to $N_{\mathrm{targ}}$ in the ``observed'' planet catalog and the ratio of the expected number of detected planets to $N_{\mathrm{targ}}$ in the simulated catalog, and take the maximum over all the bins included in the current simulation).  

\begin{figure}
\centering
\includegraphics[scale=0.43]{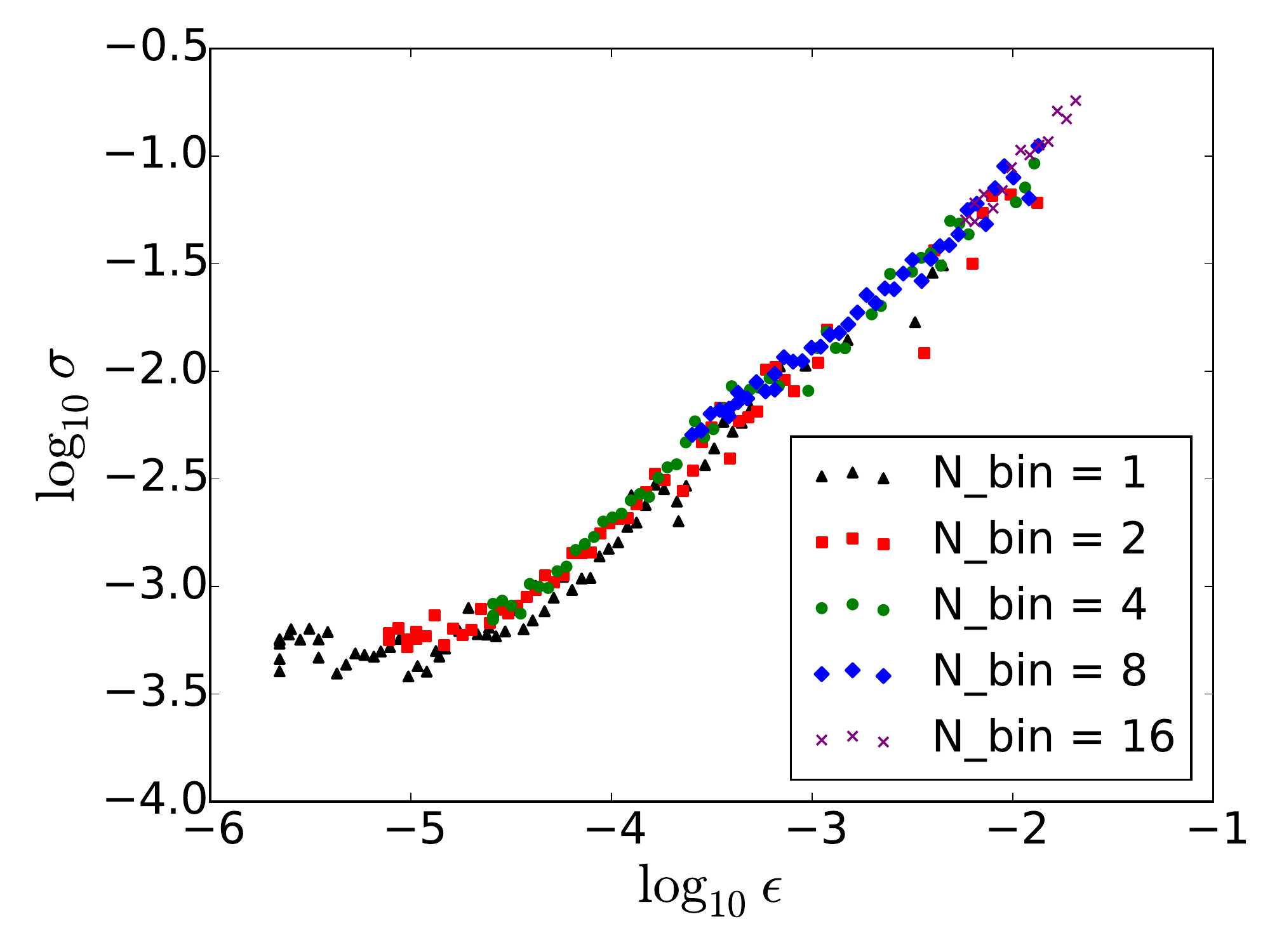}
\caption{Standard deviation ($\sigma$) of the inferred planet occurrence rate as a function of the final distance tolerance ($\epsilon$) for different choices of the number of occurrence rate parameters, $N_{\rm{bin}}$, with $N_{\star} = 40,000$ and a ``true'' occurrence rate $f = 0.015$.  Each simulation starts from the top right and decreases in $\sigma$ towards the bottom left over the course of the run.  Our implementation of ABC works well for inferring up to 4 parameters simultaneously, but modifications would be required to efficiently perform inference with $\ge8$ parameters simultaneously.
}
 \label{eps-sig-nbin}
\end{figure}

We perform tests inferring values of $f_{r,p}$ for 1, 2, 4, 8, 16 and 32 size-period bins simultaneously.  
We find that the ABC algorithm successfully converged around the true occurrence rates for the 1, 2, and 4 bin tests.
However, it became very inefficient for the 8, 16, and 32 bin tests even before achieving an $\epsilon$ that is acceptably small (see Fig. \ref{eps-sig-nbin}), i.e., $\epsilon_{\mathrm{slow}}$ was significantly larger than the $\epsilon$ needed to realize the most precise possible ABC estimates for the $f_{r,p}$'s.

\section{Application to Kepler Data}
\label{secApplication}
Having validated that our algorithm and implementations of ABC-PMC and SysSim can accurately characterize planet occurrence rates, we proceed to apply this framework to a recent Kepler planet catalog.  
The catalog of target star properties is described in \S\ref{secStarProperties}.  
For observed planet candidate catalog properties, we adopt the Q1-Q16 planet candidate catalog obtained from the NASA Exoplanet Archive in March 2017 \citep{MCT+2014,RBM+2014}.
We find $N_{\mathrm{p}} = 3380$ planet candidates associated with our target stars and having estimated planet radii ($R_{p}$) in the range $0.5-16 R_{\oplus}$ and orbital periods ($P$) in the range $0.5-320$ days.  For this catalog, the detection efficiency curve we apply is a $\Gamma$ distribution CDF with $\alpha = 4.65$ and $\beta = 0.98$, which is the appropriate detection efficiency given the completeness properties of the Q1-Q16 data release.
This input catalog has since been updated to reflect new information since \citet{CCB+2015}.
It is worth noting, however, that the Q1-Q16 data 
release has not had a systematic search across the full 
parameter space considered in this study because of issues which are discussed in detail in \citet{BCM+2015}.

\subsection{Comparison of ABC and Inverse Detection Efficiency Methods}
\label{secComparison}
\begin{figure*}
\centering
\includegraphics[scale=0.5]{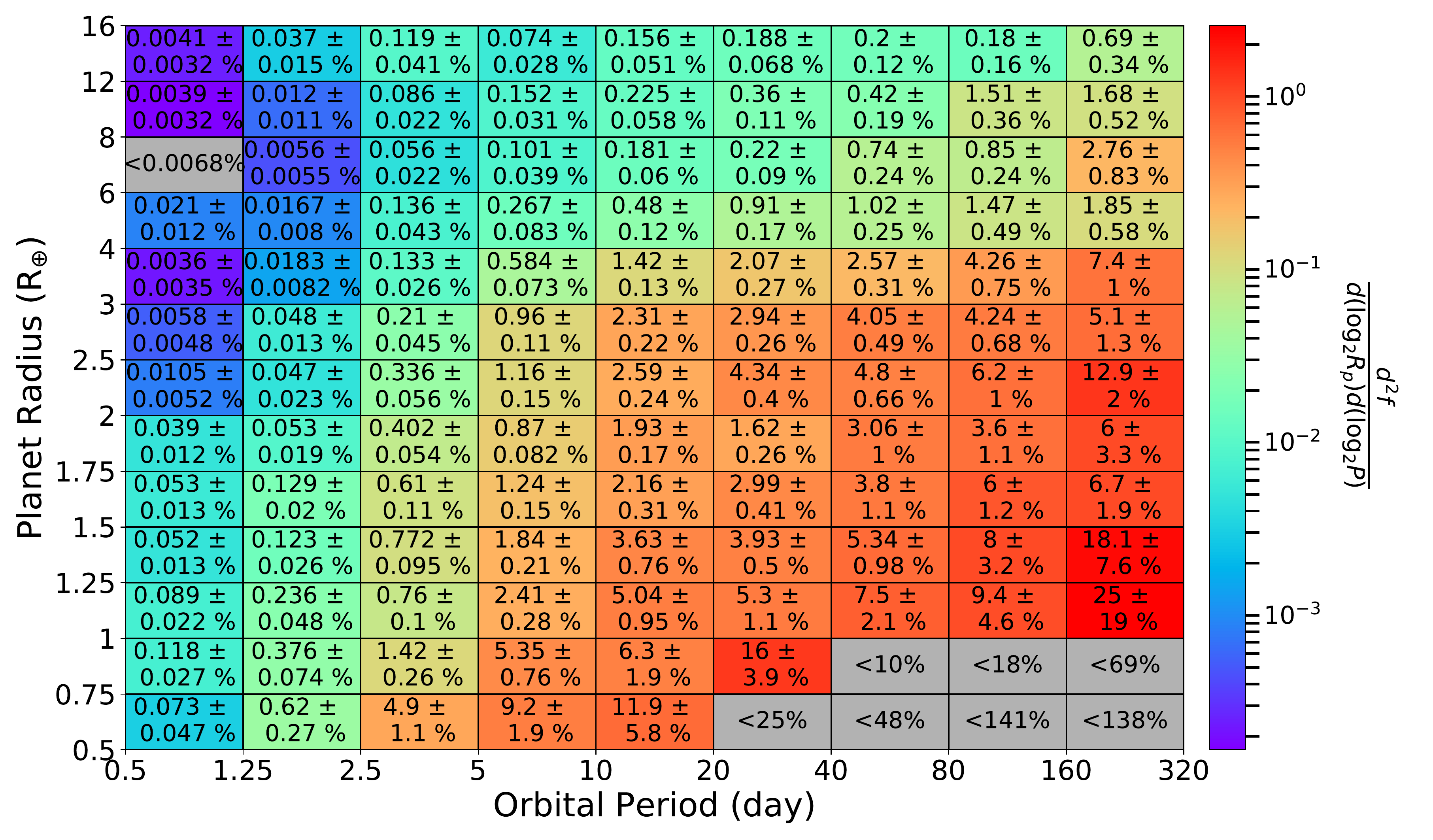}
\caption{The ABC estimated occurrence rate for the Q1-Q16 planet candidates orbiting FGK stars using the \citet{CCB+2015} Gamma CDF curve for the detection efficiency.  
The numerical values of the occurrence rates are stated as percentage (i.e. $10^{-2}$). 
The color coding of each cell is based on $(d^{2}f)/[d (\log_2{R_{p}})~d(\log_2{P})]$, which provides an occurrence rate normalized to the width of the bin and therefore is not dependent on choice of grid density.  
Cells colored gray have estimated upper limits for the occurrence rate.
Note that the bin sizes are not constant. }
\label{q1q16-rates}
\end{figure*}

First, we perform separate simulations to characterize the planet occurrence rate for each planet size-period bin individually.  
For each test, the target $\epsilon$ was set at an unrealistically small $10^{-10}$ and the maximum number of generations was set to $200$, so that ABC would continue to run until triggering one of the termination heuristics that recognize when further computation will not result in significantly reducing $\epsilon_g$ of the final generation (see \S\ref{secABCPMC}).  
We find that the ABC posterior distributions for all occurrence rates are unimodal.  
For the vast majority of bins, the ABC posterior appears roughly Gaussian, with the exception of bins with zero to a few planets detected.  
We summarize our results, by reporting the posterior mean of each planet occurrence rate and the 68.27\% credible interval based on the 15.87\% and 84.13\% percentiles in Table \ref{tab:occ_rates}.  In Figure \ref{q1q16-rates} we report the maximum of the difference between the 84.13\% percentile and the mean and the difference between the mean and the 15.87\% percentile. 

\startlongtable
\begin{deluxetable*}{rrrrr}
\tablewidth{0pt}
\tablecaption{Occurrence Rates
\label{tab:occ_rates}}
\tablehead{
\colhead{Period}&
\colhead{Radius}&
\colhead{ABC}&
\colhead{Bayesian}&
\colhead{IDEM}\\
\colhead{(days)}&
\colhead{($R_{\oplus}$)}&
\colhead{(Full Model)}&
\colhead{(Simplified Model)}&
\colhead{}}

\startdata
$\phn\phn 0.50-\phn\phn 1.25$&$\phn0.50-\phn0.75$&$7.3^{+4.7}_{-2.9}\times10^{-4}$&$7.4^{+3.2}_{-3.2}\times10^{-4}$&$(6.6\pm3.3)\times10^{-4}$\\
$\phn\phn 0.50-\phn\phn 1.25$&$\phn0.75-\phn1.00$&$1.18^{+0.22}_{-0.27}\times10^{-3}$&$1.20^{+0.24}_{-0.24}\times10^{-3}$&$(1.44\pm0.29)\times10^{-3}$\\
$\phn\phn 0.50-\phn\phn 1.25$&$\phn1.00-\phn1.25$&$8.9^{+2.2}_{-1.4}\times10^{-4}$&$9.3^{+1.7}_{-1.7}\times10^{-4}$&$(1.12\pm0.21)\times10^{-3}$\\
$\phn\phn 0.50-\phn\phn 1.25$&$\phn1.25-\phn1.50$&$5.19^{+0.96}_{-1.30}\times10^{-4}$&$5.1^{+1.2}_{-1.2}\times10^{-4}$&$(6.0\pm1.4)\times10^{-4}$\\
$\phn\phn 0.50-\phn\phn 1.25$&$\phn1.50-\phn1.75$&$5.3^{+1.3}_{-1.1}\times10^{-4}$&$5.1^{+1.1}_{-1.1}\times10^{-4}$&$(5.4\pm1.2)\times10^{-4}$\\
$\phn\phn 0.50-\phn\phn 1.25$&$\phn1.75-\phn2.00$&$3.89^{+1.24}_{-0.78}\times10^{-4}$&$3.75^{+0.96}_{-0.96}\times10^{-4}$&$(4.5\pm1.2)\times10^{-4}$\\
$\phn\phn 0.50-\phn\phn 1.25$&$\phn2.00-\phn2.50$&$1.05^{+0.52}_{-0.45}\times10^{-4}$&$9.8^{+4.7}_{-4.7}\times10^{-5}$&$(8.2\pm4.7)\times10^{-5}$\\
$\phn\phn 0.50-\phn\phn 1.25$&$\phn2.50-\phn3.00$&$5.8^{+4.8}_{-2.7}\times10^{-5}$&$7.3^{+4.0}_{-4.0}\times10^{-5}$&$(3.7\pm2.6)\times10^{-5}$\\
$\phn\phn 0.50-\phn\phn 1.25$&$\phn3.00-\phn4.00$&$3.6^{+3.5}_{-2.2}\times10^{-5}$&$4.8^{+3.1}_{-3.1}\times10^{-5}$&$(2.7\pm2.7)\times10^{-5}$\\
$\phn\phn 0.50-\phn\phn 1.25$&$\phn4.00-\phn6.00$&$2.06^{+0.55}_{-1.20}\times10^{-4}$&$9.6^{+4.6}_{-4.6}\times10^{-5}$&$(7.8\pm4.5)\times10^{-5}$\\
$\phn\phn 0.50-\phn\phn 1.25$&$\phn6.00-\phn8.00$&$<6.8\times10^{-5}$&$<2.4\times10^{-5}$&N/A\\
$\phn\phn 0.50-\phn\phn 1.25$&$\phn8.00-12.00$&$3.9^{+3.2}_{-2.3}\times10^{-5}$&$4.8^{+3.1}_{-3.1}\times10^{-5}$&$(2.8\pm2.8)\times10^{-5}$\\
$\phn\phn 0.50-\phn\phn 1.25$&$12.00-16.00$&$4.1^{+3.2}_{-2.7}\times10^{-5}$&$4.8^{+3.1}_{-3.1}\times10^{-5}$&$(3.0\pm3.0)\times10^{-5}$\\
\hline
$\phn\phn 1.25-\phn\phn 2.50$&$\phn0.50-\phn0.75$&$6.2^{+2.7}_{-1.8}\times10^{-3}$&$5.9^{+1.7}_{-1.7}\times10^{-3}$&$(1.25\pm0.38)\times10^{-2}$\\
$\phn\phn 1.25-\phn\phn 2.50$&$\phn0.75-\phn1.00$&$3.76^{+0.63}_{-0.74}\times10^{-3}$&$3.75^{+0.67}_{-0.67}\times10^{-3}$&$(4.90\pm0.89)\times10^{-3}$\\
$\phn\phn 1.25-\phn\phn 2.50$&$\phn1.00-\phn1.25$&$2.36^{+0.48}_{-0.38}\times10^{-3}$&$2.19^{+0.37}_{-0.37}\times10^{-3}$&$(2.47\pm0.43)\times10^{-3}$\\
$\phn\phn 1.25-\phn\phn 2.50$&$\phn1.25-\phn1.50$&$1.23^{+0.15}_{-0.26}\times10^{-3}$&$1.16^{+0.24}_{-0.24}\times10^{-3}$&$(1.30\pm0.28)\times10^{-3}$\\
$\phn\phn 1.25-\phn\phn 2.50$&$\phn1.50-\phn1.75$&$1.29^{+0.20}_{-0.19}\times10^{-3}$&$1.24^{+0.24}_{-0.24}\times10^{-3}$&$(1.39\pm0.27)\times10^{-3}$\\
$\phn\phn 1.25-\phn\phn 2.50$&$\phn1.75-\phn2.00$&$5.3^{+1.9}_{-1.2}\times10^{-4}$&$5.7^{+1.6}_{-1.6}\times10^{-4}$&$(6.1\pm1.8)\times10^{-4}$\\
$\phn\phn 1.25-\phn\phn 2.50$&$\phn2.00-\phn2.50$&$4.68^{+2.30}_{-0.96}\times10^{-4}$&$4.7^{+1.4}_{-1.4}\times10^{-4}$&$(4.2\pm1.3)\times10^{-4}$\\
$\phn\phn 1.25-\phn\phn 2.50$&$\phn2.50-\phn3.00$&$4.8^{+1.3}_{-1.2}\times10^{-4}$&$4.6^{+1.4}_{-1.4}\times10^{-4}$&$(5.0\pm1.6)\times10^{-4}$\\
$\phn\phn 1.25-\phn\phn 2.50$&$\phn3.00-\phn4.00$&$1.83^{+0.82}_{-0.82}\times10^{-4}$&$2.08^{+0.90}_{-0.90}\times10^{-4}$&$(2.0\pm1.0)\times10^{-4}$\\
$\phn\phn 1.25-\phn\phn 2.50$&$\phn4.00-\phn6.00$&$1.67^{+0.76}_{-0.80}\times10^{-4}$&$1.66^{+0.79}_{-0.79}\times10^{-4}$&$(1.27\pm0.74)\times10^{-4}$\\
$\phn\phn 1.25-\phn\phn 2.50$&$\phn6.00-\phn8.00$&$5.6^{+5.5}_{-3.6}\times10^{-5}$&$8.3^{+5.4}_{-5.3}\times10^{-5}$&$(3.4\pm3.4)\times10^{-5}$\\
$\phn\phn 1.25-\phn\phn 2.50$&$\phn8.00-12.00$&$1.15^{+1.12}_{-0.46}\times10^{-4}$&$1.24^{+0.68}_{-0.68}\times10^{-4}$&$(8.0\pm5.6)\times10^{-5}$\\
$\phn\phn 1.25-\phn\phn 2.50$&$12.00-16.00$&$3.7^{+1.5}_{-1.2}\times10^{-4}$&$3.3^{+1.1}_{-1.1}\times10^{-4}$&$(3.0\pm1.1)\times10^{-4}$\\
\hline
$\phn\phn 2.50-\phn\phn 5.00$&$\phn0.50-\phn0.75$&$4.94^{+1.07}_{-0.95}\times10^{-2}$&$4.60^{+0.82}_{-0.82}\times10^{-2}$&$(5.7\pm1.0)\times10^{-2}$\\
$\phn\phn 2.50-\phn\phn 5.00$&$\phn0.75-\phn1.00$&$1.42^{+0.25}_{-0.26}\times10^{-2}$&$1.30^{+0.19}_{-0.19}\times10^{-2}$&$(1.48\pm0.22)\times10^{-2}$\\
$\phn\phn 2.50-\phn\phn 5.00$&$\phn1.00-\phn1.25$&$7.60^{+1.01}_{-0.74}\times10^{-3}$&$7.17^{+0.96}_{-0.96}\times10^{-3}$&$(7.7\pm1.0)\times10^{-3}$\\
$\phn\phn 2.50-\phn\phn 5.00$&$\phn1.25-\phn1.50$&$7.72^{+0.93}_{-0.95}\times10^{-3}$&$7.00^{+0.79}_{-0.79}\times10^{-3}$&$(7.59\pm0.86)\times10^{-3}$\\
$\phn\phn 2.50-\phn\phn 5.00$&$\phn1.50-\phn1.75$&$6.12^{+1.09}_{-0.82}\times10^{-3}$&$5.85^{+0.67}_{-0.67}\times10^{-3}$&$(6.33\pm0.73)\times10^{-3}$\\
$\phn\phn 2.50-\phn\phn 5.00$&$\phn1.75-\phn2.00$&$4.02^{+0.54}_{-0.45}\times10^{-3}$&$3.97^{+0.53}_{-0.53}\times10^{-3}$&$(4.44\pm0.60)\times10^{-3}$\\
$\phn\phn 2.50-\phn\phn 5.00$&$\phn2.00-\phn2.50$&$3.36^{+0.56}_{-0.45}\times10^{-3}$&$3.11^{+0.46}_{-0.46}\times10^{-3}$&$(3.31\pm0.50)\times10^{-3}$\\
$\phn\phn 2.50-\phn\phn 5.00$&$\phn2.50-\phn3.00$&$2.10^{+0.45}_{-0.37}\times10^{-3}$&$2.15^{+0.38}_{-0.38}\times10^{-3}$&$(2.38\pm0.43)\times10^{-3}$\\
$\phn\phn 2.50-\phn\phn 5.00$&$\phn3.00-\phn4.00$&$1.33^{+0.26}_{-0.23}\times10^{-3}$&$1.32^{+0.29}_{-0.29}\times10^{-3}$&$(1.25\pm0.29)\times10^{-3}$\\
$\phn\phn 2.50-\phn\phn 5.00$&$\phn4.00-\phn6.00$&$1.36^{+0.43}_{-0.33}\times10^{-3}$&$1.38^{+0.30}_{-0.30}\times10^{-3}$&$(1.36\pm0.30)\times10^{-3}$\\
$\phn\phn 2.50-\phn\phn 5.00$&$\phn6.00-\phn8.00$&$5.6^{+2.2}_{-1.2}\times10^{-4}$&$5.9^{+1.9}_{-1.9}\times10^{-4}$&$(5.5\pm1.9)\times10^{-4}$\\
$\phn\phn 2.50-\phn\phn 5.00$&$\phn8.00-12.00$&$8.6^{+2.2}_{-2.0}\times10^{-4}$&$8.5^{+2.3}_{-2.3}\times10^{-4}$&$(8.9\pm2.6)\times10^{-4}$\\
$\phn\phn 2.50-\phn\phn 5.00$&$12.00-16.00$&$1.19^{+0.41}_{-0.19}\times10^{-3}$&$1.25^{+0.28}_{-0.28}\times10^{-3}$&$(1.11\pm0.26)\times10^{-3}$\\
\hline
$\phn\phn 5.00-\phn10.00$&$\phn0.50-\phn0.75$&$9.2^{+1.9}_{-1.9}\times10^{-2}$&$8.6^{+2.0}_{-2.0}\times10^{-2}$&$(1.15\pm0.28)\times10^{-1}$\\
$\phn\phn 5.00-\phn10.00$&$\phn0.75-\phn1.00$&$5.35^{+0.76}_{-0.60}\times10^{-2}$&$4.91^{+0.61}_{-0.61}\times10^{-2}$&$(4.87\pm0.61)\times10^{-2}$\\
$\phn\phn 5.00-\phn10.00$&$\phn1.00-\phn1.25$&$2.41^{+0.22}_{-0.28}\times10^{-2}$&$2.17^{+0.25}_{-0.25}\times10^{-2}$&$(2.09\pm0.24)\times10^{-2}$\\
$\phn\phn 5.00-\phn10.00$&$\phn1.25-\phn1.50$&$1.84^{+0.21}_{-0.18}\times10^{-2}$&$1.68^{+0.17}_{-0.17}\times10^{-2}$&$(1.74\pm0.18)\times10^{-2}$\\
$\phn\phn 5.00-\phn10.00$&$\phn1.50-\phn1.75$&$1.24^{+0.13}_{-0.15}\times10^{-2}$&$1.14^{+0.12}_{-0.12}\times10^{-2}$&$(1.17\pm0.13)\times10^{-2}$\\
$\phn\phn 5.00-\phn10.00$&$\phn1.75-\phn2.00$&$8.70^{+0.82}_{-0.69}\times10^{-3}$&$8.29^{+1.00}_{-1.00}\times10^{-3}$&$(8.7\pm1.1)\times10^{-3}$\\
$\phn\phn 5.00-\phn10.00$&$\phn2.00-\phn2.50$&$1.16^{+0.14}_{-0.15}\times10^{-2}$&$1.04^{+0.11}_{-0.11}\times10^{-2}$&$(1.16\pm0.12)\times10^{-2}$\\
$\phn\phn 5.00-\phn10.00$&$\phn2.50-\phn3.00$&$9.6^{+1.1}_{-1.1}\times10^{-3}$&$9.20^{+1.00}_{-1.00}\times10^{-3}$&$(9.7\pm1.1)\times10^{-3}$\\
$\phn\phn 5.00-\phn10.00$&$\phn3.00-\phn4.00$&$5.84^{+0.64}_{-0.73}\times10^{-3}$&$5.50^{+0.76}_{-0.76}\times10^{-3}$&$(5.70\pm0.80)\times10^{-3}$\\
$\phn\phn 5.00-\phn10.00$&$\phn4.00-\phn6.00$&$2.67^{+0.83}_{-0.52}\times10^{-3}$&$2.72^{+0.53}_{-0.53}\times10^{-3}$&$(2.65\pm0.53)\times10^{-3}$\\
$\phn\phn 5.00-\phn10.00$&$\phn6.00-\phn8.00$&$1.01^{+0.39}_{-0.32}\times10^{-3}$&$1.04^{+0.32}_{-0.32}\times10^{-3}$&$(9.6\pm3.2)\times10^{-4}$\\
$\phn\phn 5.00-\phn10.00$&$\phn8.00-12.00$&$1.52^{+0.31}_{-0.29}\times10^{-3}$&$1.56^{+0.40}_{-0.40}\times10^{-3}$&$(1.38\pm0.37)\times10^{-3}$\\
$\phn\phn 5.00-\phn10.00$&$12.00-16.00$&$7.4^{+2.2}_{-2.8}\times10^{-4}$&$8.3^{+2.9}_{-2.9}\times10^{-4}$&$(6.7\pm2.5)\times10^{-4}$\\
\hline
$\phn10.00-\phn20.00$&$\phn0.50-\phn0.75$&$1.19^{+0.58}_{-0.45}\times10^{-1}$&$1.14^{+0.42}_{-0.42}\times10^{-1}$&$(8.3\pm3.4)\times10^{-2}$\\
$\phn10.00-\phn20.00$&$\phn0.75-\phn1.00$&$6.3^{+1.9}_{-1.3}\times10^{-2}$&$5.8^{+1.1}_{-1.1}\times10^{-2}$&$(6.8\pm1.3)\times10^{-2}$\\
$\phn10.00-\phn20.00$&$\phn1.00-\phn1.25$&$5.04^{+0.64}_{-0.95}\times10^{-2}$&$4.12^{+0.53}_{-0.53}\times10^{-2}$&$(3.77\pm0.49)\times10^{-2}$\\
$\phn10.00-\phn20.00$&$\phn1.25-\phn1.50$&$3.63^{+0.76}_{-0.31}\times10^{-2}$&$3.15^{+0.33}_{-0.33}\times10^{-2}$&$(3.22\pm0.34)\times10^{-2}$\\
$\phn10.00-\phn20.00$&$\phn1.50-\phn1.75$&$2.16^{+0.31}_{-0.24}\times10^{-2}$&$1.93^{+0.22}_{-0.22}\times10^{-2}$&$(1.87\pm0.21)\times10^{-2}$\\
$\phn10.00-\phn20.00$&$\phn1.75-\phn2.00$&$1.93^{+0.15}_{-0.17}\times10^{-2}$&$1.66^{+0.18}_{-0.18}\times10^{-2}$&$(1.68\pm0.19)\times10^{-2}$\\
$\phn10.00-\phn20.00$&$\phn2.00-\phn2.50$&$2.59^{+0.24}_{-0.16}\times10^{-2}$&$2.40^{+0.21}_{-0.21}\times10^{-2}$&$(2.64\pm0.23)\times10^{-2}$\\
$\phn10.00-\phn20.00$&$\phn2.50-\phn3.00$&$2.31^{+0.22}_{-0.18}\times10^{-2}$&$2.11^{+0.19}_{-0.19}\times10^{-2}$&$(2.21\pm0.20)\times10^{-2}$\\
$\phn10.00-\phn20.00$&$\phn3.00-\phn4.00$&$1.42^{+0.11}_{-0.13}\times10^{-2}$&$1.38^{+0.15}_{-0.15}\times10^{-2}$&$(1.38\pm0.15)\times10^{-2}$\\
$\phn10.00-\phn20.00$&$\phn4.00-\phn6.00$&$4.78^{+1.19}_{-0.61}\times10^{-3}$&$4.84^{+0.89}_{-0.89}\times10^{-3}$&$(4.55\pm0.86)\times10^{-3}$\\
$\phn10.00-\phn20.00$&$\phn6.00-\phn8.00$&$1.81^{+0.60}_{-0.52}\times10^{-3}$&$1.66^{+0.52}_{-0.52}\times10^{-3}$&$(1.46\pm0.49)\times10^{-3}$\\
$\phn10.00-\phn20.00$&$\phn8.00-12.00$&$2.25^{+0.49}_{-0.58}\times10^{-3}$&$2.33^{+0.61}_{-0.61}\times10^{-3}$&$(2.51\pm0.70)\times10^{-3}$\\
$\phn10.00-\phn20.00$&$12.00-16.00$&$1.56^{+0.51}_{-0.50}\times10^{-3}$&$1.66^{+0.52}_{-0.52}\times10^{-3}$&$(1.47\pm0.49)\times10^{-3}$\\
\hline
$\phn20.00-\phn40.00$&$\phn0.50-\phn0.75$&$<2.5\times10^{-1}$&$<1.1\times10^{-1}$&N/A\\
$\phn20.00-\phn40.00$&$\phn0.75-\phn1.00$&$1.60^{+0.39}_{-0.29}\times10^{-1}$&$1.38^{+0.30}_{-0.30}\times10^{-1}$&$(1.43\pm0.32)\times10^{-1}$\\
$\phn20.00-\phn40.00$&$\phn1.00-\phn1.25$&$5.30^{+1.05}_{-0.87}\times10^{-2}$&$4.53^{+0.88}_{-0.88}\times10^{-2}$&$(3.88\pm0.78)\times10^{-2}$\\
$\phn20.00-\phn40.00$&$\phn1.25-\phn1.50$&$3.93^{+0.50}_{-0.45}\times10^{-2}$&$3.33^{+0.51}_{-0.51}\times10^{-2}$&$(3.56\pm0.56)\times10^{-2}$\\
$\phn20.00-\phn40.00$&$\phn1.50-\phn1.75$&$2.99^{+0.37}_{-0.41}\times10^{-2}$&$2.47^{+0.35}_{-0.35}\times10^{-2}$&$(2.57\pm0.37)\times10^{-2}$\\
$\phn20.00-\phn40.00$&$\phn1.75-\phn2.00$&$1.62^{+0.21}_{-0.26}\times10^{-2}$&$1.36^{+0.23}_{-0.23}\times10^{-2}$&$(1.49\pm0.25)\times10^{-2}$\\
$\phn20.00-\phn40.00$&$\phn2.00-\phn2.50$&$4.34^{+0.34}_{-0.40}\times10^{-2}$&$3.62^{+0.34}_{-0.34}\times10^{-2}$&$(3.80\pm0.36)\times10^{-2}$\\
$\phn20.00-\phn40.00$&$\phn2.50-\phn3.00$&$2.94^{+0.26}_{-0.25}\times10^{-2}$&$2.68^{+0.28}_{-0.28}\times10^{-2}$&$(2.68\pm0.28)\times10^{-2}$\\
$\phn20.00-\phn40.00$&$\phn3.00-\phn4.00$&$2.07^{+0.27}_{-0.17}\times10^{-2}$&$1.90^{+0.23}_{-0.23}\times10^{-2}$&$(1.90\pm0.23)\times10^{-2}$\\
$\phn20.00-\phn40.00$&$\phn4.00-\phn6.00$&$9.1^{+1.7}_{-1.2}\times10^{-3}$&$8.4^{+1.5}_{-1.5}\times10^{-3}$&$(7.8\pm1.4)\times10^{-3}$\\
$\phn20.00-\phn40.00$&$\phn6.00-\phn8.00$&$2.20^{+0.90}_{-0.68}\times10^{-3}$&$2.15^{+0.74}_{-0.74}\times10^{-3}$&$(2.11\pm0.80)\times10^{-3}$\\
$\phn20.00-\phn40.00$&$\phn8.00-12.00$&$3.6^{+1.1}_{-1.0}\times10^{-3}$&$3.75^{+0.99}_{-0.99}\times10^{-3}$&$(3.49\pm0.97)\times10^{-3}$\\
$\phn20.00-\phn40.00$&$12.00-16.00$&$1.88^{+0.68}_{-0.63}\times10^{-3}$&$1.88^{+0.69}_{-0.69}\times10^{-3}$&$(1.80\pm0.73)\times10^{-3}$\\
\hline
$\phn40.00-\phn80.00$&$\phn0.50-\phn0.75$&$<4.8\times10^{-1}$&$<1.9\times10^{-1}$&N/A\\
$\phn40.00-\phn80.00$&$\phn0.75-\phn1.00$&$<1.0\times10^{-1}$&$<4.3\times10^{-2}$&N/A\\
$\phn40.00-\phn80.00$&$\phn1.00-\phn1.25$&$7.5^{+2.1}_{-1.9}\times10^{-2}$&$5.9^{+1.7}_{-1.7}\times10^{-2}$&$(6.4\pm1.9)\times10^{-2}$\\
$\phn40.00-\phn80.00$&$\phn1.25-\phn1.50$&$5.34^{+0.77}_{-0.98}\times10^{-2}$&$3.84^{+0.85}_{-0.85}\times10^{-2}$&$(3.73\pm0.86)\times10^{-2}$\\
$\phn40.00-\phn80.00$&$\phn1.50-\phn1.75$&$3.80^{+1.06}_{-0.81}\times10^{-2}$&$3.00^{+0.56}_{-0.56}\times10^{-2}$&$(3.09\pm0.59)\times10^{-2}$\\
$\phn40.00-\phn80.00$&$\phn1.75-\phn2.00$&$3.06^{+1.00}_{-0.53}\times10^{-2}$&$2.29^{+0.41}_{-0.41}\times10^{-2}$&$(2.47\pm0.45)\times10^{-2}$\\
$\phn40.00-\phn80.00$&$\phn2.00-\phn2.50$&$4.80^{+0.60}_{-0.66}\times10^{-2}$&$4.00^{+0.47}_{-0.47}\times10^{-2}$&$(3.92\pm0.47)\times10^{-2}$\\
$\phn40.00-\phn80.00$&$\phn2.50-\phn3.00$&$4.05^{+0.41}_{-0.49}\times10^{-2}$&$3.25^{+0.39}_{-0.39}\times10^{-2}$&$(3.09\pm0.38)\times10^{-2}$\\
$\phn40.00-\phn80.00$&$\phn3.00-\phn4.00$&$2.57^{+0.25}_{-0.31}\times10^{-2}$&$2.26^{+0.32}_{-0.32}\times10^{-2}$&$(2.09\pm0.30)\times10^{-2}$\\
$\phn40.00-\phn80.00$&$\phn4.00-\phn6.00$&$1.02^{+0.21}_{-0.25}\times10^{-2}$&$9.7^{+2.1}_{-2.1}\times10^{-3}$&$(9.1\pm2.0)\times10^{-3}$\\
$\phn40.00-\phn80.00$&$\phn6.00-\phn8.00$&$7.4^{+2.4}_{-1.6}\times10^{-3}$&$7.0^{+1.7}_{-1.7}\times10^{-3}$&$(6.3\pm1.6)\times10^{-3}$\\
$\phn40.00-\phn80.00$&$\phn8.00-12.00$&$4.2^{+1.9}_{-1.2}\times10^{-3}$&$4.4^{+1.4}_{-1.4}\times10^{-3}$&$(4.0\pm1.3)\times10^{-3}$\\
$\phn40.00-\phn80.00$&$12.00-16.00$&$1.99^{+1.20}_{-0.76}\times10^{-3}$&$2.20^{+0.95}_{-0.95}\times10^{-3}$&$(1.84\pm0.92)\times10^{-3}$\\
\hline
$\phn80.00-160.00$&$\phn0.50-\phn0.75$&$<1.4\times10^{0}$&$<4.8\times10^{-1}$&N/A\\
$\phn80.00-160.00$&$\phn0.75-\phn1.00$&$<1.8\times10^{-1}$&$<7.3\times10^{-2}$&N/A\\
$\phn80.00-160.00$&$\phn1.00-\phn1.25$&$9.4^{+4.1}_{-4.6}\times10^{-2}$&$7.5^{+3.2}_{-3.2}\times10^{-2}$&$(4.4\pm2.2)\times10^{-2}$\\
$\phn80.00-160.00$&$\phn1.25-\phn1.50$&$8.0^{+3.2}_{-2.1}\times10^{-2}$&$5.6^{+1.7}_{-1.7}\times10^{-2}$&$(5.4\pm1.7)\times10^{-2}$\\
$\phn80.00-160.00$&$\phn1.50-\phn1.75$&$6.0^{+1.2}_{-1.2}\times10^{-2}$&$4.3^{+1.0}_{-1.0}\times10^{-2}$&$(3.76\pm0.94)\times10^{-2}$\\
$\phn80.00-160.00$&$\phn1.75-\phn2.00$&$3.58^{+1.05}_{-0.62}\times10^{-2}$&$2.69^{+0.65}_{-0.65}\times10^{-2}$&$(2.49\pm0.62)\times10^{-2}$\\
$\phn80.00-160.00$&$\phn2.00-\phn2.50$&$6.15^{+1.03}_{-0.96}\times10^{-2}$&$4.42^{+0.68}_{-0.68}\times10^{-2}$&$(4.48\pm0.70)\times10^{-2}$\\
$\phn80.00-160.00$&$\phn2.50-\phn3.00$&$4.24^{+0.68}_{-0.63}\times10^{-2}$&$3.28^{+0.52}_{-0.52}\times10^{-2}$&$(3.51\pm0.56)\times10^{-2}$\\
$\phn80.00-160.00$&$\phn3.00-\phn4.00$&$4.26^{+0.60}_{-0.75}\times10^{-2}$&$3.35^{+0.50}_{-0.50}\times10^{-2}$&$(3.22\pm0.49)\times10^{-2}$\\
$\phn80.00-160.00$&$\phn4.00-\phn6.00$&$1.47^{+0.49}_{-0.36}\times10^{-2}$&$1.30^{+0.30}_{-0.30}\times10^{-2}$&$(1.30\pm0.32)\times10^{-2}$\\
$\phn80.00-160.00$&$\phn6.00-\phn8.00$&$8.5^{+2.4}_{-2.0}\times10^{-3}$&$7.9^{+2.4}_{-2.4}\times10^{-3}$&$(7.7\pm2.4)\times10^{-3}$\\
$\phn80.00-160.00$&$\phn8.00-12.00$&$1.51^{+0.36}_{-0.22}\times10^{-2}$&$1.44^{+0.32}_{-0.32}\times10^{-2}$&$(1.44\pm0.33)\times10^{-2}$\\
$\phn80.00-160.00$&$12.00-16.00$&$1.79^{+1.55}_{-0.97}\times10^{-3}$&$2.2^{+1.2}_{-1.2}\times10^{-3}$&$(1.6\pm1.1)\times10^{-3}$\\
\hline
$160.00-320.00$&$\phn0.50-\phn0.75$&$<1.4\times10^{0}$&$<8.0\times10^{-1}$&N/A\\
$160.00-320.00$&$\phn0.75-\phn1.00$&$<6.9\times10^{-1}$&$<2.4\times10^{-1}$&N/A\\
$160.00-320.00$&$\phn1.00-\phn1.25$&$2.5^{+1.9}_{-1.3}\times10^{-1}$&$1.59^{+0.87}_{-0.86}\times10^{-1}$&$(6.2\pm4.4)\times10^{-2}$\\
$160.00-320.00$&$\phn1.25-\phn1.50$&$1.81^{+0.76}_{-0.44}\times10^{-1}$&$1.30^{+0.45}_{-0.45}\times10^{-1}$&$(1.74\pm0.66)\times10^{-1}$\\
$160.00-320.00$&$\phn1.50-\phn1.75$&$6.7^{+1.9}_{-1.7}\times10^{-2}$&$4.3^{+1.7}_{-1.7}\times10^{-2}$&$(4.7\pm2.1)\times10^{-2}$\\
$160.00-320.00$&$\phn1.75-\phn2.00$&$6.0^{+3.3}_{-1.7}\times10^{-2}$&$4.1^{+1.3}_{-1.3}\times10^{-2}$&$(3.4\pm1.1)\times10^{-2}$\\
$160.00-320.00$&$\phn2.00-\phn2.50$&$1.29^{+0.20}_{-0.18}\times10^{-1}$&$8.2^{+1.4}_{-1.4}\times10^{-2}$&$(8.7\pm1.5)\times10^{-2}$\\
$160.00-320.00$&$\phn2.50-\phn3.00$&$5.08^{+0.87}_{-1.27}\times10^{-2}$&$3.19^{+0.72}_{-0.72}\times10^{-2}$&$(3.14\pm0.74)\times10^{-2}$\\
$160.00-320.00$&$\phn3.00-\phn4.00$&$7.4^{+1.0}_{-1.0}\times10^{-2}$&$5.26^{+0.86}_{-0.86}\times10^{-2}$&$(5.59\pm0.93)\times10^{-2}$\\
$160.00-320.00$&$\phn4.00-\phn6.00$&$1.85^{+0.46}_{-0.58}\times10^{-2}$&$1.62^{+0.46}_{-0.46}\times10^{-2}$&$(1.62\pm0.49)\times10^{-2}$\\
$160.00-320.00$&$\phn6.00-\phn8.00$&$2.76^{+0.80}_{-0.83}\times10^{-2}$&$2.42^{+0.56}_{-0.56}\times10^{-2}$&$(2.41\pm0.59)\times10^{-2}$\\
$160.00-320.00$&$\phn8.00-12.00$&$1.68^{+0.52}_{-0.47}\times10^{-2}$&$1.61^{+0.46}_{-0.46}\times10^{-2}$&$(1.57\pm0.47)\times10^{-2}$\\
$160.00-320.00$&$12.00-16.00$&$6.9^{+3.4}_{-2.7}\times10^{-3}$&$6.7^{+2.9}_{-2.9}\times10^{-3}$&$(5.8\pm2.9)\times10^{-3}$\\
\enddata

\tablecomments{ABC, Poisson, and inverse detection efficiency method (IDEM) estimated occurrence rates for Q1-Q16 KOI catalog planet candidates associated with FGK stars.}
\end{deluxetable*}

Due to updates to the Kepler planet candidate catalog and stellar catalogs at the Exoplanet Archive, a direct comparison of occurrence rates to those of \citet{CCB+2015} was not possible.
Therefore, we computed occurrence rates using the IDEM \citep{CCB+2015} and our exact same catalog of target stars and planet candidates.  
We present the corresponding planet occurrence rates computed using the inverse detection efficiency method in Table \ref{tab:occ_rates}.  Details are provided in Appendix \ref{appIDEM}.
For the sake of comparison, Table \ref{tab:occ_rates} also presents rates and uncertainties calculated using the simplified Bayesian model described in Appendix \ref{appPoisson}.

\begin{figure*}
\centering
\includegraphics[scale=0.5]{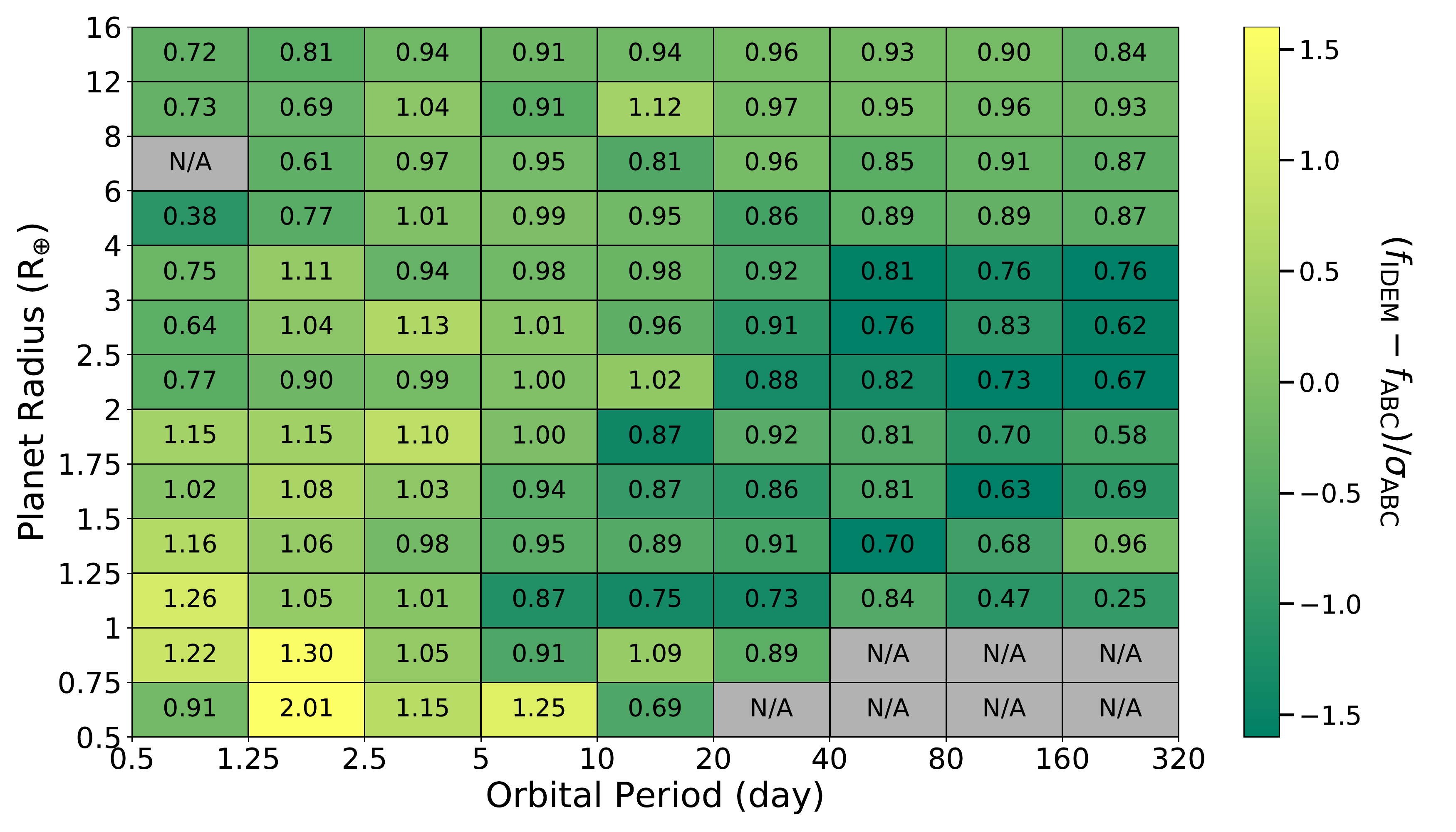}
\caption{The ratio of the planet occurrence rate inferred by the inverse detection efficiency method to the planet occurrence rate inferred the ABC method based on the Q1-Q16 candidates orbiting FGK stars. The color indicates the 
relative difference between the occurrence rates scaled by the uncertainty given by the standard deviation from the ABC method.
The two methods give similar results for easily detected planets (most bins with high signal-to-noise are within $\sim$1 standard deviations), but the inverse detection efficiency method substantially underestimates the frequency of near-Earth-size planets for orbital periods beyond $\sim~80$ days.}
\label{q1q16-ratio}
\end{figure*}

We find that the three methods yield similar planet occurrence rates and uncertainties for most of the planet radius-period bins considered.  
We show the ratio of the planet occurrence rates estimated by the IDEM to those estimated by our more rigorous ABC method in Fig. \ref{q1q16-ratio}.

The values agree very well for large planet candidates over all period ranges and planet candidates over most of the radius range with $P < 5$ days.  For these bins, there is a sufficiently large sample of planet candidates detected at high transit signal-to-noise, that the two methods give similar results, despite the noise induced by low transit signal-to-noise planet candidates. 
While the ratio of posterior means appears to differ for some bins corresponding to large planets at short orbital periods, these estimates are not not meaningfully inconsistent with each other, since these bins have few or no detections and both methods have large uncertainties.  
In contrast, small planets at short periods have occurrence rates that are inconsistent between ABC and IDEM due to IDEM not accounting for measurement uncertainties.

In summary, we find substantial differences in the estimated planet occurrence rates for planet sizes and orbital periods near the threshold of detectability (i.e. small planets at long periods), which varies slightly depending on the KOI catalog and choice of detection efficiency model.
For our baseline case, the differences are particularly notable for planets smaller than 1.5 $R_{\oplus}$ at orbital periods greater than 80 days.
Of particular interest, we show that the IDEM can underestimate the occurrence rate of planets with sizes and orbital periods similar to that of the Earth by up to a factor of $\sim~4$.  
An example of how the IDEM underestimates the planet occurrence rate for long orbital period and small planet radius bins is shown in Fig. \ref{figabc-invdet_comp}.  
Note that these estimates are based on the same data, so one would expect that the mean rates should agree to better than the width of the posterior, if both methods were providing consistent and unbiased estimates of the same quantity.  
The bias of the inverse detection efficiency method is to be expected, as described in Appendix \ref{appIDEM}.
We discuss the significant implications of this result in \S\ref{secDiscussion}.  

In the case of the simplified Bayesian model, while the occurrence rates agree well with the occurrence rates estimated using the full Bayesian model and ABC throughout much of the chosen parameter space, the simplified Bayesian model 
underestimates the occurrence rate for planet candidates near the edge of detectability (i.e. large period, small radius).   
In contrast, when estimating occurrence rates in the limit of minimal transit depth uncertainty (as done in \S\ref{secEpsilon} to show ABC posteriors closely approximate the correct posterior width) there is close agreement over all parameter space between the two models. 
As a result, the difference between the occurrence rates can be explained by the inclusion of realistic transit depth uncertainty in the ABC forward model.

\subsection{Correlations in Occurrence Rates due to Uncertainty in Planet Radii}
\label{secRadiusUncertainty}
\label{secMultiBinResults}
For the planet size-orbital period distribution model considered in this paper, the planet occurrence rates in each bin are very nearly decoupled.  
The only way that a planet with size and period in one bin can affect the occurrence rate in another bin is if the measurement error results in assigning the planet to a different size-period bin.  
Since orbital periods are measured extremely precisely, planets will almost certainly be assigned to the correct period range.  
The uncertainties in planet radius are significantly larger, both due to uncertainty in the estimated transit depth and due to the uncertainty in the host star radius.  
Our ABC approach accounts for this effect, even in simulations that characterize the occurrence rate for a single bin in planet size and orbital period.  
By performing inference on the occurrence rate for each bin separately, we do lose information about the joint ABC posterior, i.e., correlations between occurrence rates in different bins.  

To test the significance of this effect, we performed simulations that simultaneously estimate occurrence rates for multiple bins include differing planet size ranges, but a common orbital period.   
To quantify how much these uncertainties have on our occurrence rate estimates, we performed ABC simulations to estimate the joint ABC posterior for the planet occurrence rates for neighboring planet size bins (up to two on each side), each time using a single common orbital period range.  

As expected based on results from \S\ref{secNParam}, the final $\epsilon$ is not as small as when characterizing each occurrence rate separately.
This results in a modest increase in the width of the marginal ABC posterior for each occurrence rate.  
Next, we investigated the correlation between occurrence rate in neighboring planet size bins.  
We did not fine large correlations for any pair of neighboring bins, with a maximum absolute correlation coefficient of 0.4.

\section{Discussion}
\label{secDiscussion}
%
%\subsection{Context and Importance}
NASA's Kepler mission was designed to measure the occurrence rate of exoplanets, particularly small planets in the habitable zone of solar-like host stars.  
Kepler was enormously successful, finding thousands of strong planet candidates with sizes less than Neptune.  
However, since the probability of a planet both transiting and being detectable by Kepler falls off rapidly as one approaches Earth-size planets in the habitable zone of sun-like stars, Kepler identified precious few potential Earth-analogues.  
The small number of detections and rapidly changing detection efficiency for small planets necessitates careful attention be paid to statistical methodology when interpreting the Kepler results in terms of intrinsic planet populations.  
Many early investigations of the planet occurrence rates based on Kepler data relied on the simplistic IDEM which is known to be significantly biased for some of the most interesting regions of parameter space.  
Some more recent studies have attempted to improve on this by using a Bayesian framework for estimating planet occurrence rates (e.g., \citet{BCM+2015}) and incorporating hierarchical models (e.g., \citet{FHM2014}), providing significant advances in statistical rigor and power.
We build on these works by introducing Approximate Bayesian Computing to perform inference with hierarchical Bayesian models for the planet occurrence rates.  \textit{We emphasize that these calculations are done with several simplifying assumptions, particularly in terms of the input catalog used (see \S\ref{secLimitations}).}  Therefore, we regard these as preliminary estimates which improve on many previous studies, but still leave significant room for further improvement.  \textit{Substantial further research is needed to improve the accuracy and precision of occurrence rates before making important decisions based on planet occurrence rates (e.g., design of a large space mission).}

\subsection{Summary of Key Results}
%
% * ABC works
We have demonstrated that HBM and the ABC-PMC algorithm can accurately characterize planet occurrence rates.  
%
% * ABC-PMC is already practical for <= 8 model parameters
%
% * Inverse detection efficiency is significantly biased for small planets
We confirm that the inverse detection efficiency method can provide accurate estimates of the planet occurrence rate for planets with sufficiently high transit signal-to-noise and short orbital periods.  
We also show that the inverse detection efficiency method is significantly biased when applied to planets near the threshold of detection, leading to it underestimating the planet occurrence rate by a factor of up to 4 for nearly Earth-size planets with orbital periods greater than 80 days when applied to the Kepler Q1-Q16 catalog.

% * Planet occurrence rates for Kepler catalog
We applied our HBM and ABC methodology to the Kepler Q1-16 planet candidate catalog.
%
% * ? Comparison of Planet occurrence rates for two Kepler catalogs ?
We characterized the rate of planet candidates over a larger range of planet sizes and with finer detail than most previous studies.  
%
% * ? How smooth are the occurrence rates ?
We find a relatively smooth planet occurrence rate in planet size and orbital period over most of the range explored, even without assuming a smooth functional form.  However, we find a ``radius valley'' (i.e., a local minimum in the marginalized planet occurrence rate per logarithmic interval in planet size), particularly for orbital periods between 20 and 80 days.
This finding is consistent with other studies which found a radius valley based on smaller samples for which more precisely measured stellar properties were available \citep[e.g.,][]{FPH+2017,VAL+2017}.  Interestingly, we find that this feature can be identified even without the improved spectroscopic or asteroseismic constraints, albeit not as strongly as if constraints were incorporated.  The fact that three studies each find a significant radius valley, despite substantial differences in the target sample used, planet candidate catalog, and the origin of the stellar properties assumed, points to the robustness of this feature in the Kepler data.
 
We find that a joint power law in planet size and orbital period is not an adequate model when considering planets ranging from Earth-size to $\sim~2 R_\oplus$ or larger and periods 20-320 days.
Given the observed structure in the planet occurrence rate, even a broken power-law is inadequate when considering planets ranging from Earth-size to $\sim~3 R_\oplus$ or larger.  

We also provide a simplified Bayesian model for planet occurrence rates that avoids the bias inherent in the IDEM.
We recommend that future studies that want a fast method of estimating occurrence rates in the high signal-to-noise regime adopt this simplified Bayesian model, rather than continuing to use the IDEM.  Of course, a full hierarchical Bayesian model would be even better, since the simplified Bayesian model neglects measurement uncertainties.

\begin{figure}
\centering
\includegraphics[scale=0.3]{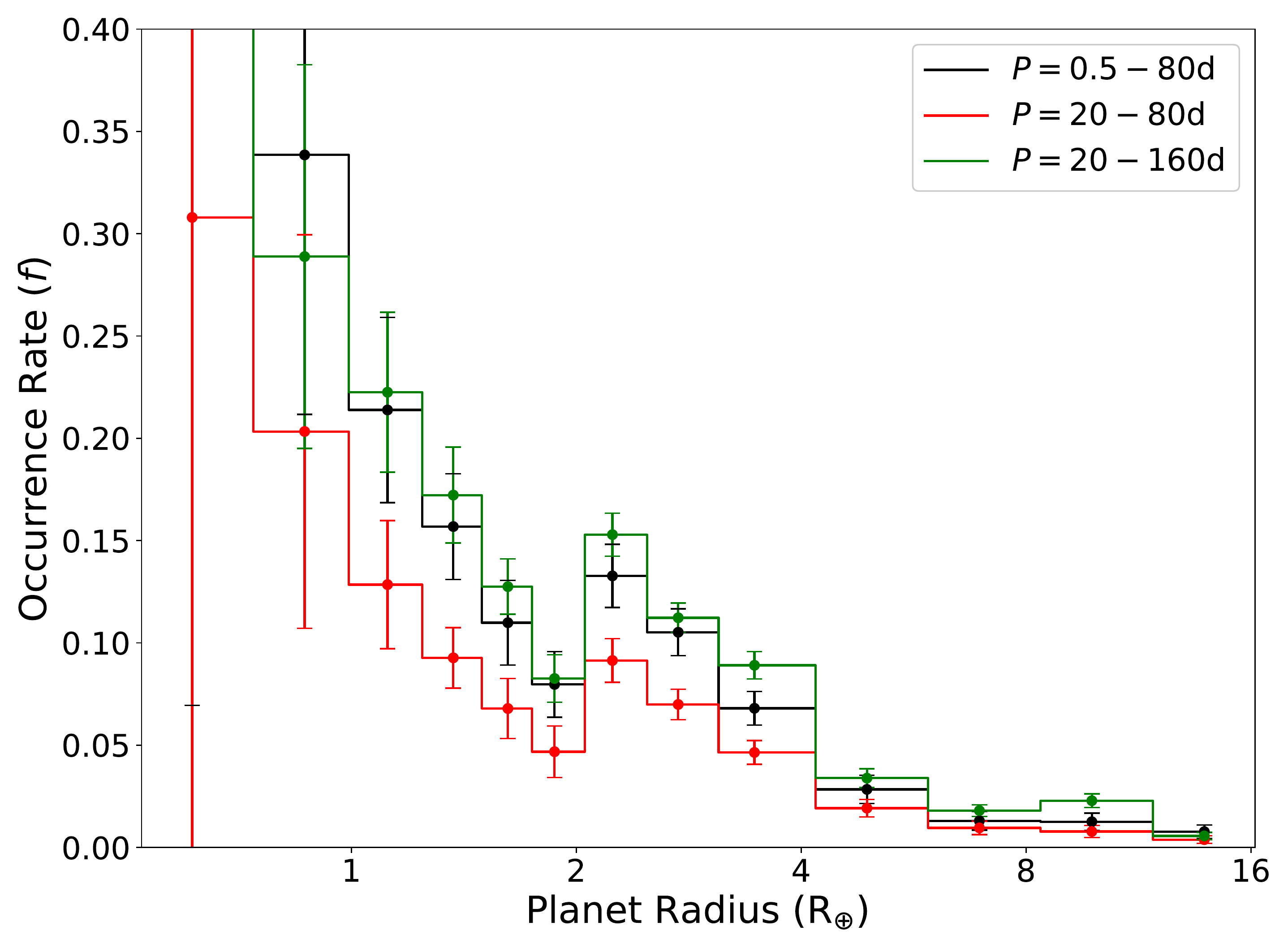}
\caption{ABC estimated occurrence rates from Fig. \ref{q1q16-rates} integrated over three different ranges of period bins to emphasize the existence of a local minimum near $1.75-2 R_{\oplus}$.
}
\label{q1q16-gap}
\end{figure}

\subsection{Limitations of Our Occurrence Rates}
\label{secLimitations}
While these results represent a significant methodological advance, several astrophysical assumptions affect this and other published occurrence rate studies.  
For example, uncertainties in the radius of host stars directly affects the uncertainty in planet radii.  
While the ABC method accounts for uncertainty in the stellar properties, we have assumed errors follow a Gaussian distribution with the reported uncertainties.  
In reality there are likely non-Gaussian measurement uncertainties and potentially systematic biases in the measured stellar properties.
While considerable effort has gone into creating and updating the stellar properties catalog, we expect that the stellar properties could be significantly improved once parallaxes are available from GAIA.  

As another example, we have assumed that all KOIs not identified as a likely false positive are strong planet candidates.  
In practice, some are likely astrophysical false positives, such as a brown dwarf or M star that transits one star in a hierarchical triple system.  
Other planet candidates may be false alarms that survived the vetting process.
Similarly, we have neglected the effects of the transit depths being diluted due to more than one star falling into the aperture used for constructing Kepler light curves.  
We adopted a single fixed distribution for the orbital eccentricity, rather than simultaneously inferring the eccentricity distribution or allowing for a dependence on planet properties.  
Based on previous sensitivity studies, we expect that this has a modest effect on our results \citep{BCM+2015}, smaller than the uncertainties due to the choice of Kepler planet candidate catalog, host stars properties and potential inaccuracy of the model for transit detection and vetting efficiency and reliability.  
In principle, our ABC framework makes it practical to incorporate these and additional effects into a HBM analysis of exoplanet populations.  
However, incorporating such effects is beyond the scope of this study.  

\subsection{Implications for $\eta_{\oplus}$}
\label{secEtaEarth}
Our results have significant implications for estimates of $\eta_{\oplus}$, the rate of nearly Earth-size planets in the habitable zone of solar-type stars.  
Many definitions of $\eta_{\oplus}$ can be found in the literature.  
Following the convention of the NASA Exoplanet Program Analysis Group, Study Analysis Group 13's report on Exoplanet Occurrence Rates and Distributions, we compute the occurrence rate for planets with radii $1-1.5 R_\oplus$ and orbital periods 237-320 days \citep{SAG13}.
Applying our HBM and ABC method to the Kepler Q1-Q16 catalog, we find a rate of $0.41^{+0.29}_{-0.12}$ per star, larger than the $0.11\pm0.08$ that would be estimated using the same data set and assumptions with the inverse detection efficiency method.  This underscores the importance of quantifying the accuracy and biases of various statistical methodologies for computing planet occurrence rates.

As mentioned in \S\ref{secLimitations}, there are multiple complications that likely affect the inferred rate of small planets and are not reflected in the uncertainty estimates above.  In particular, the above estimates are for the rate of planet candidates, as reported in the Q1-Q16 catalog.  Any astrophysical false positives or statistical false alarms that managed to survive the vetting process would contribute to increasing our inferred rate of planet candidates.  For most of the planet sizes and orbital periods considered, the Kepler pipeline has a low rate of false alarms and good sensitivity to reject astrophysical false positives.  However, based on the DR25 completeness products the rate of contamination from false positives and false alarms is greater in the range of periods and sizes affecting estimates of $\eta_{\oplus}$, since the vetting process is more sensitive for planet candidates with larger transit signal-to-noise and shorter orbital periods.  
Subsequent to the calculations reported in this paper, the Kepler team released additional data products that could be used to characterize the performance of the Kepler pipeline \citep{TCH2018}.  We recommend that future research to consider how to incorporate these data products into occurrence rate studies in a statistically valid way.

Of course, the largest uncertainty in the rate of small planets in the habitable zone is the size of the habitable zone.  
Therefore, we also report a differential occurrence rate per star, $(d^2 f)/[d(\log_2 P)~d(\log_2 R_{p})] = 1.6^{+1.2}_{-0.5}$ for $R_p=1-1.5 R_{\oplus},P=237-320\mathrm{d}$, so other researchers can easily scale our result to their preferred ranges of planet sizes and orbital periods. If one were to extrapolate to a wider range of orbital periods and assume that the rate density remains constant, then one would arrive at an occurrence rate of $1.0^{+0.7}_{-0.3}$ per star for $R_p=1-1.5 R_{\oplus},P=237-500\mathrm{d}$ which covers an 
optimistic range for the habitable zone.  Of course, one should recognize the limitations of such an extrapolation, as physical processes could cause the planet occurrence rate to vary significantly over this range.

\subsection{Future Prospects for Occurrence Rate Studies}
\label{secFutureRates}
Future studies will further improve on planet occurrence rates based on the Kepler data set.  
While there is clearly room for improvements in the population modeling (e.g., \S\ref{secLimitations}), we anticipate that these will have relatively minor effects compared to the anticipated improvements in the catalogs of planet candidates and star properties.  
In particular, we are currently incorporating results from the improved pipeline detection efficiency released for the DR25 KOI catalog as provided by the Kepler Science Team.  The DR25 KOI catalog is based on the complete data set, a final Kepler pipeline and an automated vetting process \citep{TCH2018}.  We have shown that the inferred planet occurrence rates are sensitive to the assumed detection efficiency.  Therefore, we refrain from presenting results based on the DR25 catalog until the pipeline detection efficiency is well-characterized and implemented in our model.  
We also anticipate significant improvements due to improved characterization of host stars once GAIA parallaxes become available.  
We hope that this study will make it practical for future planet occurrence rate studies that incorporating these improved data sets to apply a rigorous hierarchical Bayesian model.
In particular, we provide open source codes (at GitHub, see $\S\ref{secDistance}$ to facilitate the use of ABC in future studies, including estimating planet occurrence rates, as well as any number of problems that involve population inference.

\subsection{Future Prospects for Characterizing Population of Exoplanet Architectures}
\label{secFutureSysSim}
While numerous studies have attempted to characterize the rate of individual planets, the frequency of various planetary system architectures is an even more powerful constraint on planet formation theories.  
Perhaps the most important contribution of this study is to provide a practical framework for performing statistical inference on the population of planetary \emph{systems}.  

While the geometric transit probability for a single planet is relatively simple ($P \sim \sin^{-1}(R_{\star}/a)$) for circular orbits with $a > R_{\star}$, the transit probability for a system with multiple planets is much more complicated.  
Additional selection effects (e.g., the detection efficiency and window function) further complicate comparing the predictions of theoretical models to the Kepler results.  
Since these considerations are extremely difficult to incorporate in occurrence rate analyses of exoplanet catalogs, previous studies have neglected the effects of multiple planet systems.  
While previous studies could only focus on overall occurrence rates based on planetary property distributions, SysSim and ABC provide a means for rigorously addressing questions like: ``How common are systems with at least one pair of planets with orbital periods within 2\% of a 2:1 orbital resonance?''
This ability to characterize the intrinsic rate of various planetary architectures will enable a wide range of theoretical studies to better understanding of planet formation.

To address such questions, one must model the joint distribution for planetary system properties, e.g., allowing for potential correlations in the orbital period and sizes of planets in the same planetary system.  
With SysSim and the ABC approach, one has access to the true physical parameters for each planetary system in the simulated catalog.  
Thus, one can compute the transit and detection probabilities for any combination of the planets in a system.  
In the most straightforward approach, one can construct simulated catalogs based on which planets would transit for a single specified viewing geometry and their detection efficiencies.  
SysSim can also includes advanced functionality to increase the computational efficiency of ABC when characterizing planetary architectures.
For many summary statistics related to multiple planet systems (e.g., number of detected planets, ratios of their orbital periods), SysSim can include weights, analogous to how this study weighted each potentially detectable planet by a detection probability that was averaged over all viewing geometries.
The required sky-averaged probabilities can be calculated efficiently using nearly algebraic
formulae that integrate over all viewing directions for any subset of the planets in each system using CORBITS \citep{BR2016}.

Realizing this potential will require further methodological research.
For example, when using multiple qualitatively different summary statistics, it will be important to explore the impact of various choices for distance functions.  
An adequate model for the population of exoplanetary systems may require several parameters, motivating further research in improving the efficiency of the sequential importance sampler used by ABC-PMC.  

\subsection{Future Prospects for Characterizing Exoplanet Populations Including Follow-Up Observations}
\label{secFutureFollowUp}
In \S\ref{secHBM}, we highlighted the difficulties of characterizing a population of planetary systems when most of the planets are undetected.  
Another application of ABC is in cases where practical considerations make it unrealistic to even write down a likelihood function.
For example, in principle, a study characterizing the planet occurrence rate around a given type of star should account for the target selection criteria.  
That would require a likelihood that integrates over all possible target stars, dramatically increasing the computational complexity.  
The choice of target stars for the Kepler planet search depended on a complex merit function that was based on the estimates of stellar properties prior to launch which are now out-of-date.
Further complicating matters, whether a star was targeted depended not only on its own properties, but also the properties of other stars that were being considered for targeting.  
Without conditional independence, the integrals over properties of each potential target can not be separated.

As another example, if one wanted to include constraints from ground-based follow-up observations, one should account for the process of choosing targets worthy of follow-up observations, the whims of time allocation committees selecting successful proposals, the limitations of weather, how observers change their plans based on preliminary data analysis, etc.  
While writing down a likelihood that accounts for such effects is not realistic, it can be feasible to implement a forward model that reasonably simulates such effects.  
Our ABC approach has the potential to enable future studies that account for target selection and even follow-up observations.  
We anticipate that this could be particularly important for future analyses of planet populations identified by NASA's Transiting Exoplanet Survey Satellite (TESS) before being targeted by ground-based Doppler follow-up.

\acknowledgments
We thank the entire Kepler team for the many years of work that has proven so successful and was critical to this study.  
D.C.H, E.B.F., R.C.M and D.R. acknowledge support from NASA Origins of Solar Systems grant \# NNX14AI76G and helpful discussions with Keir Ashby.
E.B.F and R.C.M. acknowledge support from NASA Kepler Participating Scientist Program Cycle II grant \# NNX14AN76G,
D.C.H, E.B.F. acknowledge support from the Penn State Eberly College of Science and Department of Astronomy \& Astrophysics, the Center for Exoplanets and Habitable Worlds and the Center for Astrostatistics.  
The citations in this paper have made use of NASA's Astrophysics Data System Bibliographic Services.  
This research has made use of the NASA Exoplanet Archive, which is operated by the California Institute of Technology, under contract with the National Aeronautics and Space Administration under the Exoplanet Exploration Program.
We acknowledge the Institute for CyberScience (\url{http://ics.psu.edu/}) at The Pennsylvania State University, including the CyberLAMP cluster supported by NSF grant MRI-1626251, for providing advanced computing resources and services that have contributed to the research results reported in this paper.
This material was based upon work partially supported by the National Science Foundation under Grant DMS-1127914 to the Statistical and Applied Mathematical Sciences Institute (SAMSI). Any opinions, findings, and conclusions or recommendations expressed in this material are those of the author(s) and do not necessarily reflect the views of the National Science Foundation.
This study benefited from the 2013 SAMSI workshop on Modern Statistical and Computational Methods for Analysis of Kepler Data, the 2016/2017 Program on Statistical, Mathematical and Computational Methods for Astronomy, and their associated working groups.

\appendix
\section{Inverse Detection Efficiency Method}
\label{appIDEM}
The inverse detection efficiency method (IDEM) attempts to calculate weights from observed planet candidates in order to ``correct'' for the planets which weren't observed by Kepler.

Following closely to the methodology of \citet{CCB+2015}, the IDEM used in this study is as follows:
\begin{enumerate}
\item Calculate the geometric transit probability of each planet candidate $i$ about its host star: $p_{g,i} = R_{\star}/a$
\item Calculate the probability each planet candidate $i$ has of being detected if it transits around each target star $j$ in the stellar catalog (given the planet has the same orbital period but a uniformly chosen random impact parameter between 0 and 1) using the chosen detection efficiency curve: $p_{d,i,j}$.
\item Sum $p_{d,i,j}$ over all target stars and divide by the total number of target stars to get the fraction of stars around which a given planet candidate $i$ would have been detected: $p_{d,i}/N_{\mathrm{targ}} = \left(\sum_{j=1}^{N_{\mathrm{targ}}} p_{d,i,j}\right)/N_{\mathrm{targ}}$.
\item Calculate a weight $C_{i} = 1/(p_{g,i}p_{d,i})$ that is often interpreted as an estimate of the number of planets of the same size and period as planet $i$ in the Kepler survey, attempting to account for incompleteness due to geometric transit probability and detection efficiency.
\item Summing the weights $C_i$ for planets with periods and radii within the limits of a bin and dividing by the total number of target stars gives an estimate of the occurrence rate within that bin: $f_{r,p} = \left(\sum_{i~\mathrm{in~bin~r,p}} C_{i}\right)/N_{\mathrm{targ}}$
\end{enumerate}

For the associated uncertainty, we report the inverse of the Fischer information (curvature of the log likelihood) for the rate parameter under a Poisson distribution, where the rate parameter is the estimated planet occurrence rate and the number of planet detections within a bin $N_{r,p}$ from the Kepler data.
\begin{equation}
\sigma_{f_{r,p}} = \frac{f_{r,p}}{\sqrt{N_{r,p}}}
\end{equation}

A significant shortcoming of this technique for estimating occurrence rates lies with the fact that it does not properly marginalize over the uncertainty in the transit parameters.   Inevitably, measurement errors cause the estimated transit properties (e.g., transit depth, duration) to deviate from the true transit parameters.  If the estimated transit SNR is smaller (larger) than the true transit SNR, then the planet is less (more) likely to be detected, i.e., measurement noise reduces (increases) the measured signal.  For a planet with transit SNR near the threshold of detection, the detection probability changes rapidly as a function of transit SNR, so there can be a large difference between the detection probability for the true transit SNR and the estimated transit SNR.  This difference impacts the occurrence rates estimated by IDEM in two ways.  First, it affects which planets are included in the catalog of planet detections.  The smallest detected planets are likely to have their sizes over-estimated than under-estimated, since this dramatically increases their detection probability.  
Second, it affects the values of $p_{d,i}$ and hence $C_i$ which are assigned to the planets.  For the smallest detected planets, the estimates for the planet size, transit SNR and detection probability are all biased to be larger than the true values, causing IDEM to systematically overestimate $p_{d,i}$ and thus systematically underestimate $C_i$, the occurrence rate, $f_{r,p}$, and the uncertainty in the occurrence rate, $\sigma_{f_{r,p}}$.

Because the detection probability changes strongly and non-linearly as a function of planet size, this decrease in the estimated occurrence rate from smaller planets is not fully compensated for by an increase in the estimated occurrence rate larger from larger planets (whose true SNR may be overestimated).
For larger planets, the transit SNR is larger than the threshold of detection, and the the detection probability is changing much more slowly, so the magnitude of these effects are dramatically reduced. 
We demonstrate the resulting bias in planet occurrence rates by analyzing simulated data sets with the IDEM in Figure \ref{figabc-invdet_comp}.
This effect is strongest for occurrence rate estimates that include planets near the threshold of detection, such as nearly Earth-size planets near the habitable zone (see Fig.\ \ref{q1q16-ratio}).  It is important to note that
this bias can often be partially corrected with 
a few modifications as we have done with a simplified Bayesian model in Appendix \ref{appPoisson}.  See also the appendix of \citet{FHM2014}.

\section{Simplified Bayesian Model for Planet Occurrence Rates}
\label{appPoisson}
Here we present a simplified Bayesian model for estimating planet occurrence rates.  
We assume that the log period and log radius of planets are drawn from a Poisson point process with constant rate ($(d^2f)/[d(\log_2 P)~d(\log_2 R_{p})]$) over each bin defined by a range of orbital periods and planet sizes.  
Then the number of planets around each star within the bin is distributed as a Poisson random variable with a rate parameter ($f_{r,p} = \left.(d^2f)/[d(\log_2 P)~d(\log_2 R_p)]\right|_{r,p} \Delta_r \Delta_p$) proportional to the area of the bin ($\Delta_r \Delta_p$).

If we could observe $N_{\mathrm{targ}}$ targets and detect all planets, then the sum of the number of planets in the bin ($N_{\mathrm{all},r,p}$) over all target stars would also a Poisson random variable with rate parameter $f_{r,p} N_{\mathrm{targ}}$.
If we let the prior for the rate parameter ($f_{r,p}$) be a Gamma distribution with shape parameter $\alpha_0$ and rate parameter (also known as the inverse scale parameter) $\beta_0$, then the posterior distribution for $f_{r,p}$ will be a Gamma distribution with shape parameter $\alpha_0+N_{\mathrm{all},r,p}$ and rate parameter $\beta_0+N_{\mathrm{targ}}$.  

In practice, we do not detect all planets due to a combination of geometric transit probability and detection efficiency.  
Instead, we are only sensitive to planets around a subset of the the target stars, primarily dictated by the orientation of the orbital plane and the photometric precision for each star.  
If we assume that, for each target star (indexed by $j$) and bin in planet size and orbital period, the combined probability of detecting its planets ($p_{r,p,j}$) is nearly constant, then the effective number of stars searched ($N_{\mathrm{ESS},r,p}$, where $r$ and $p$ indicate the range of planet sizes and orbital periods) is the sum of $N_{\mathrm targ}$ Bernoulli random variables with success probabilities $p_{r,p,j}$.  
In this model, the posterior distribution for $f_{r,p}$ conditioned on detecting $N_{r,p}$ planets will again be a Gamma distribution,
\begin{equation}
\label{eqnSimpleBayesPosterior}
p(f_{r,p} | N_{r,p}, N_{\mathrm{ESS},r,p} ) \sim \mathrm{Gamma}(\alpha_0+N_{r,p}, \beta_0+N_{\mathrm{ESS},r,p}).
\end{equation}  
Since $N_{\mathrm{ESS},r,p}$ is itself a random variable, one would like to marginalize over $N_{\mathrm{ESS},r,p}$, but the resulting posterior for $f_{r,p}$ is not analytic and is computationally expensive.  

We can approximate the posterior for $f_{r,p}$ by substituting the expected value, $E\left[N_{\mathrm{ESS},r,p}\right]$ for the random variable  $N_{\mathrm{ESS},r,p}$ in Eqn.\ \ref{eqnSimpleBayesPosterior}.  
We estimate the expected value of the effective number of stars searched via Monte Carlo.  
For each star and range of planet sizes and orbital periods, we draw $N_{\mathrm{samp}}=100$ planets (uniformly in $\log P$ and $\log R_{p}$ within the bin limits) and estimate the average value of $p_{r,p,j} \simeq (1/N_{\mathrm{samp}}) \sum_{i=1}^{N_{\mathrm{samp}}} p_{\mathrm{geo},i,j} p_{\mathrm{det},i,j}$ via Monte Carlo, where $p_{\mathrm{geo},i,j}$ 
and $p_{\mathrm{det},i,j}$ are the geometric transit probability and detection probability for the $i$th draw of planet parameters.  
The expected value for the effective number of stars searched is simply the sum of these probabilities over all target stars,
$E\left[N_{\mathrm{ESS},r,p}\right] = \sum_{j=1}^{N_{\mathrm{targ}}} p_{r,p,j}$.
Then we approximate the posterior for the intrinsic occurrence rate of planets in each range of planet sizes and orbital periods by
\begin{equation}
\label{eqnSimpleBayesPosteriorApprox}
p(f_{r,p} | N_{r,p}, E\left[N_{\mathrm{ESS},r,p}\right] ) \sim \mathrm{Gamma}(\alpha_0+N_{r,p}, \beta_0+E\left[N_{\mathrm{ESS},r,p}\right]).
\end{equation}  
To help build intuition for this result is useful to consider the mean and standard deviation of the gamma distribution posterior,
\begin{equation}
\mu_{f_{r,p}} = \frac{\alpha_0+N_{r,p}}{\beta_0+E\left[N_{\mathrm{ESS},r,p}\right]}
\end{equation}
\begin{equation}
\sigma_{f_{r,p}} = \frac{\mu_{f_{r,p}}}{\sqrt{\beta_0+E\left[N_{\mathrm{ESS},r,p}\right]}}.
\end{equation}
If $\alpha_0$ and $\beta_0$ are much less than $N_{r,p}$ and $E\left[N_{\mathrm{ESS},r,p}\right]$, then the posterior mean for the occurrence rate is essentially the ratio of the number of planets detected to the effective sample size.  
Similarly, the width of the posterior behaves as expected for a Monte Carlo estimate of a mean, in the limit of a large effective sample size.
The primary difference from IDEM is that our simplified model provides a principled definition of the effective sample size that differs from that assumed arbitrarily by IDEM.

In Table \ref{tab:occ_rates} we report occurrence rates estimated from this simplified Bayesian model, using the gamma distribution of Eqn.\ \ref{eqnSimpleBayesPosteriorApprox} in App. \ref{appPoisson} and  $\alpha_0=\beta_0=1$, which corresponds to a prior $p(f_{r,p}) = \exp(-f_{r,p})$.  
For radius-period bins where either of these quantities is zero or of order unity, then the observations provide limited information, so the resulting posterior for $f_{r,p}$ will be sensitive to the choice of prior.  
Fortunately, for most bins, the results are insensitive to the choice of prior, since both $N_{r,p}$ and $E\left[N_{\mathrm{ESS},r,p}\right]$ are much larger than unity.

The occurrence rates computed using the simplified Bayesian model agree well with the occurrence rates estimated using the full Bayesian model and ABC throughout most of the parameter space we consider.  
This represents a significant improvement upon the IDEM for which we find significant differences occur across the full range of periods for planets smaller than Neptune.  
Therefore, we recommend that future studies that want a fast method of estimating occurrence rates in the high signal-to-noise regime of parameter space adopt this simplified Bayesian model, rather than continuing to use the IDEM.

Upon more detailed inspection, one can observe that at long orbital periods, the simplified Bayesian model appears to slightly underestimate the planet occurrence rate of small planets relative to ABC.
The source of the difference can be traced to how the two models decides which occurrence rate bin a planet contributes to.  The simplified Bayesian model always assigns the planet to a period-radius bin based on the planet candidate's observed parameters as reported in the planet candidate catalog.
In contrast, the ABC method allows for a planet to be assigned to a given been to be due to a planet whose true parameters would place it in a different bin.  This strict placement of planets by the simplified Bayesian model results in an overestimated $E\left[N_{\mathrm{ESS},r,p}\right]$ when the  transit measurement noise has a significant chance of causing a planet to be assigned to a different period-radius bin.  The overestimate of $E\left[N_{\mathrm{ESS},r,p}\right]$  leads to an underestimate of the occurrence rate for those bins.  This demonstrates the importance of applying a hierarchical Bayesian model, regardless of whether computed via ABC or more traditional methods, to accurately infer the occurrence rate of small, long-period planets.

\section{Probabilistic Catalogs}
\label{appProbabilisticCatalog}
In the present study, we generate a single observed catalog, where each planet is either detected or not detected.  SysSim also has the capability to generate a probabilistic catalog, where each potentially detectable planet is included along with a weight proportional to its detection probability.  This approach could be useful for accelerating calculations or for comparing the frequency of planetary systems that are intrinsically rare or rarely detected.  
In the probabilistic catalog approach, we average over all possible observer orientations by using the sky-averaged geometric transit probability for each planet, $R_\star/[a(1-e^2)]$.  
As in the previous case, the geometric transit probability is multiplied by an updated transit detection probability.  
In this approach, we draw a single value of the impact parameter from a uniform distribution between zero and unity for the purposes of computing the transit detection probability.  
Each potentially detectable planet (indexed by $l$) is added to the observed planet catalog, along with its measured transit parameters and a weight ($w_l = p_{\mathrm{comb,l}}$) corresponding to the product of its sky-averaged geometric transit probability, the window function for it transiting at least three transits, and the detection efficiency conditioned on the planet transiting at least three times.  
The expected number of detected planets within each planet size-period divided by the number of target stars is given by
\begin{equation}
E[n_{r,p}] / N_{\mathrm{targ}} \simeq \sum_{l~\mathrm{in~bin}\, {r,p}} w_l /N_{\mathrm{targ}}
% I(\hat{R}_{p,l} \in [R_{p,r}, R_{p,r+1}]) I(P_{l} \in [P_{p}, P_{p+1}])
\end{equation}
Comparing this approach to the single viewing geometry approach, using a single viewing geometry has the advantage of simplicity and being readily extensible to incorporating more complex summary statistics (e.g., abundance of multiple detected transiting planets).  
Averaging over viewing geometries has the advantage of enabling a much more accurate estimate of the expected number of detected planets for a given $N_{\mathrm{targ}}$ which can then be chosen to be significantly less than the number of stars targeted by Kepler. 

However, one should not simply replace the number of planets in a period-radius bin with the expected value for the number of planets within that bin, as the process of averaging over the sky results in the expected value having less variance than the actual number of planets in the same bin for any given observer.  Further, the sky-averaging feature could complicate the interpretation of correlations of planets within one planetary system.  Therefore, considerable care is necessary before applying SysSim's sky averaging inside an ABC calculation.
A mathematical derivation necessary to motivate the choice of summary statistics and distance function is beyond the scope of this paper.  
This appendix merely documents this feature of SysSim.  

\bibliographystyle{aasjournal}
\bibliography{references}

\end{document}